\documentclass[14pt, preprint2]{aastex6}
\usepackage{bm, graphicx, subfigure, amsmath, morefloats, mathtools, amsfonts, amsbsy}
\hypersetup{allcolors = [rgb]{0., 0., 0.5}} 

\usepackage{longtable}
\usepackage{color}
\newcommand{\project}[1]{\textsl{#1}}

\newcommand{\apogee}{\project{\textsc{apogee}}}

\newcommand{\kpc}{\ensuremath{\mathrm{kpc}}}
\newcommand{\teff}{\mbox{$T_{\rm eff}$}}

\newcommand{\galr}{\mbox{$R_{\rm gal}$}}
\newcommand{\vhelio}{\mbox{$V_{\rm helio}$}}

\newcommand{\kms}{\mbox{$\rm kms^{-1}$}}
\newcommand{\kmskpc}{\mbox{$\rm kms^{-1}kpc$}}
\newcommand{\feh}{\mbox{$\rm [Fe/H]$}}
\newcommand{\mgfe}{\mbox{$\rm [Mg/Fe]$}}

\newcommand{\xfee}{\mbox{$\rm [X/Fe]_{error}$}}

\newcommand{\mg}{\mbox{$\rm [Mg/Fe]$}}

\newcommand{\logg}{\mbox{$\log g$}}

\newcommand{\xone}{\ensuremath{x_n}}
\newcommand{\xtwo}{\ensuremath{x_{n'}}}


\begin{document}

\title{The homogeneity of the star forming environment of the Milky Way disk over time}

\author{Melissa~K.~Ness\altaffilmark{1,2}, 
Adam J. Wheeler\altaffilmark{1}, 
Kevin McKinnon\altaffilmark{3}, 
Danny Horta\altaffilmark{4}, 
Andrew R. Casey\altaffilmark{5}, 
Emily C. Cunningham\altaffilmark{2},
Adrian~M~Price-Whelan\altaffilmark{2}}

\altaffiltext{1}{Department of Astronomy, Columbia University, Pupin Physics Laboratories, New York, NY 10027, USA}\
\altaffiltext{2}{Center for Computational Astrophysics, Flatiron Institute, 162 Fifth Avenue, New York, NY 10010, USA}
\altaffiltext{3}{Department of Astronomy and Astrophysics, University of California, Santa Cruz, CA 95064, USA}\\
\altaffiltext{4}{Astrophysics Research Institute, Liverpool John Moores University, 146 Brownlow Hill, Liverpool L3 5RF, UK}\\
\altaffiltext{5}{School of Physics \& Astronomy, Monash University, Wellington Rd, Clayton 3800, Victoria, Australia}\\

\email{mkness@gmail.com}

\begin{abstract} 
Stellar abundances and ages afford the means to link chemical enrichment to galactic formation. In the Milky Way, individual element abundances show tight correlations with age, which vary in slope across ([Fe/H]-[$\alpha$/Fe]). Here, we step from characterising abundances as measures of age, to understanding how abundances trace properties of stellar birth-environment in the disk over time. Using measurements from $\sim$27,000 APOGEE stars (R=22,500, SNR$>$200), we build simple local linear models to predict a sample of elements (X = Si, O, Ca, Ti, Ni, Al, Mn, Cr) using (Fe, Mg) abundances alone, as fiducial tracers of supernovae production channels. Given [Fe/H] and [Mg/H], we predict these elements, [X/H], to about double the uncertainty of their measurements. The intrinsic dispersion, after subtracting measurement errors in quadrature is $\approx 0.015-0.04$~dex.  The residuals of the prediction (measurement $-$ model) for each element demonstrate that each element has an individual link to birth properties at fixed (Fe, Mg).  Residuals from primarily massive-star supernovae (i.e. Si, O, Al) partially correlate with guiding radius. Residuals from primarily supernovae Ia (i.e. Mn, Ni) partially correlate with age. A fraction of the intrinsic scatter that persists at fixed (Fe, Mg), however, after accounting for correlations, does not appear to further discriminate between birth properties that can be traced with present-day measurements. Presumably, this is because the residuals are also, in part, a measure of the typical (in)-homogeneity of the disk's stellar birth environments, previously inferred only using open-cluster systems. Our study implies at fixed birth radius and time, there is a median scatter of $\approx 0.01-0.015$ dex in elements generated in supernovae sources. 
\end{abstract}

\section{Introduction}

Chemical abundances from surveys like APOGEE \citep{Majewski2017}, GALAH \citep{Buder2018} and LAMOST \citep{Lamost2012}, combined with a reference set of ages from Kepler \citep{Kepler2010} provide the means to link chemical enrichment to galactic formation over time. Temporally, the disk formed inside-out \citep[e.g.][]{Frankel2019}: we observe there to be older stars in the inner galaxy and younger stars in the outer regions. Chemically, the disk shows a negative metallicity gradient across Galactic radius, that has presumably weakened over time due to radial migration \citep[e.g.][]{Roskar2008, Frankel2018, Minchev2018}.

For stars in the disk, studies of individual element distributions, [X/Fe], have revealed that at fixed metallicity, [Fe/H], there are very small intrinsic scatters ($\approx 0.03$ dex) around weak age-individual abundance relations  \citep[e.g.][]{Nissen2015, Bedell2018, Ness2019, Hayden2020, Casa2021, ER2021}. The age-abundance relations vary in their slopes at different values in the chemical plane defined by Supernovae type Ia (SNIa) and Supernovae type II (SNII) sources (Fe, Mg); although the intrinsic scatter in the disk population is comparably small across chemical cells (or bins) in the [Fe/H]-[Mg/Fe] plane \citep{Lu2021}. Previously, we have addressed the question of how individual abundances at fixed [Fe/H] and stellar ages can link to birth radius \citep[e.g.][]{Ratcliffe2021, Ness2019}. Here, we turn this question around to ask; at fixed birth radius and age, what is the intrinsic scatter in individual elements? That is, what is the abundance scatter set by the time and place of the star forming environment? 

To address this question we first measure how well a set of elements that are produced by supernovae sources are predicted by two fiducial supernovae sources. For the disk, a two-process model defined by (Fe, Mg) can well describe the smooth individual abundance variation across the disk  \citep[][]{Weinberg2018, Emily2021, Weinberg2021}. Here we use Fe and Mg as fiducial SNIa and supernovae SNII sources, respectively, to quantify how well we can predict an ensemble of eight other supernovae produced elements (Si, O, Ca, Ti, Ni, Al, Mn, Cr), given these two elements alone. We then examine the residual scatter around these predictions and estimate how much of this scatter is \textit{not} correlated with astrophysical sources. That is, how much scatter is intrinsic to the birth place of stars in the disk.  Section 2 introduces our data, Section 3, our method, Section 4 our results, and in Sections 5 and 6 we include a brief discussion and summary of future prospects.

\section{Data}

We assemble a high fidelity sample of APOGEE DR16 red giant stars with ten abundance measurements from the ASPCAP pipeline \citep{GP2016}. We use X = Fe, Mg, Si, O, Ca, Ti, Ni, Mn, Cr, Al (we take the reported abundances with respect to Fe and convert them to [X/H] by $+$[Fe/H])\footnote{we test using both [X/H] and [X/Fe] and for local linear regression and all the results and inferences are the same}. These ten elements are understood to be produced in supernovae channels, specifically via massive star and white-dwarf explosions.  \citep[e.g.][]{K2020}. We perform the following quality cuts to obtain our sample of $\sim$27,000 stars:\\

\noindent $\teff = 4500-5500$~K \\
$\logg = 1.5-3.5$~dex \\
$\feh > -1$~dex\\
$\xfee < 0.1$~dex \\
SNR $> 200$ \\
$|\vhelio| < 50~\kms$ \\
Flags \texttt{ASPCAPBAD} and \texttt{ROTATION} not set\\

Most of these stars are located near the Sun, with a Galactic radius of  $\galr = 8.6 \pm 1~\kpc$, but range from $\galr = 2-15~\kpc$, in total. The stars span all ages, from stars as young as $\ge 0.3$~Gyr, with a mean distribution of age $= 6.2 \pm 3.6$~Gyr. At Signal to Noise (SNR) $>$ 200 these ten elements are measured to a precision of between $0.01-0.04$ dex, on average. We restrict our sample to heliocentric velocities of $-50~\kms < \vhelio < 50~\kms$ based on an initial assessment of how precisely two fiducial channels (Fe, Mg) can predict the other abundances [X/H]. We find that abundance residuals (measured abundance-predicted abundance) that we calculate show strong correlations with velocity. This is presumably not related to the star, but nuisance imprints in the spectra that are a significant fraction of the residual amplitude, from sky emission and tellurics and diffuse interstellar bands \citep[][and McKinnon in preparation]{Holtz2018}. 

We examine the correlations between individual element abundance residuals and orbital actions and ages. The residuals are calculated as the difference between the ASPCAP measurements and the model's prediction. The ages we use are derived from spectra using data-driven modeling and the Kepler reference objects \citep{Lu2021}. These have individual uncertainties of $\sigma_{\mbox{age}} \approx 3$~Gyr. The orbital actions and spatial coordinates are from the ASTRO-NN catalogue \citep{Leung2019}. Additionally, we use the open cluster members from the OCCAM catalogue  of open clusters \citep{occam} to assess the residuals in sites of known inter and intra birth origin. For examining the open cluster stars we relax our SNR limit to SNR $>$ 100 and remove our radial velocity restriction. This increases the number of field stars that we  work with to $\sim$80,000 and increases our open cluster sample by a factor of two (to $\sim$100 open cluster stars in nine clusters). We note that our less restrictive cuts  negligibly change our overall results and conclusions for the full ensemble analyses. However, using the higher fidelity sample of stars is a more precise examination of the data, with a lesser contribution from nuisance signals i.e. artifacts that correlate with radial velocity. 

\section{Method}

We want to determine, given stellar evolutionary state and two abundances (Fe, Mg), how well we can predict other elements generated via supernovae channels. To do this, we use linear regression. A local linear regression model is like basic regression \citep[][]{llr}, but an individual model is built for each object on a subset of data i.e. a local model. Each local model is constructed by defining a neighborhood around each object in some data space \citep[e.g. see][for an example of this using Kepler data]{Sayeed2021}. This approach takes on the following steps for our $\mbox{N}=27,000$ stars, where the parameters that we select as predictors are $\vec{Y}$ = (\teff, \logg, [Mg/H], [Fe/H]) and our goal is to predict eight abundances X = (Si, O, Ca, Ti, Ni, Al, Mn, Cr): \\

\begin{figure*}[]
\centering
\includegraphics[scale=0.6]{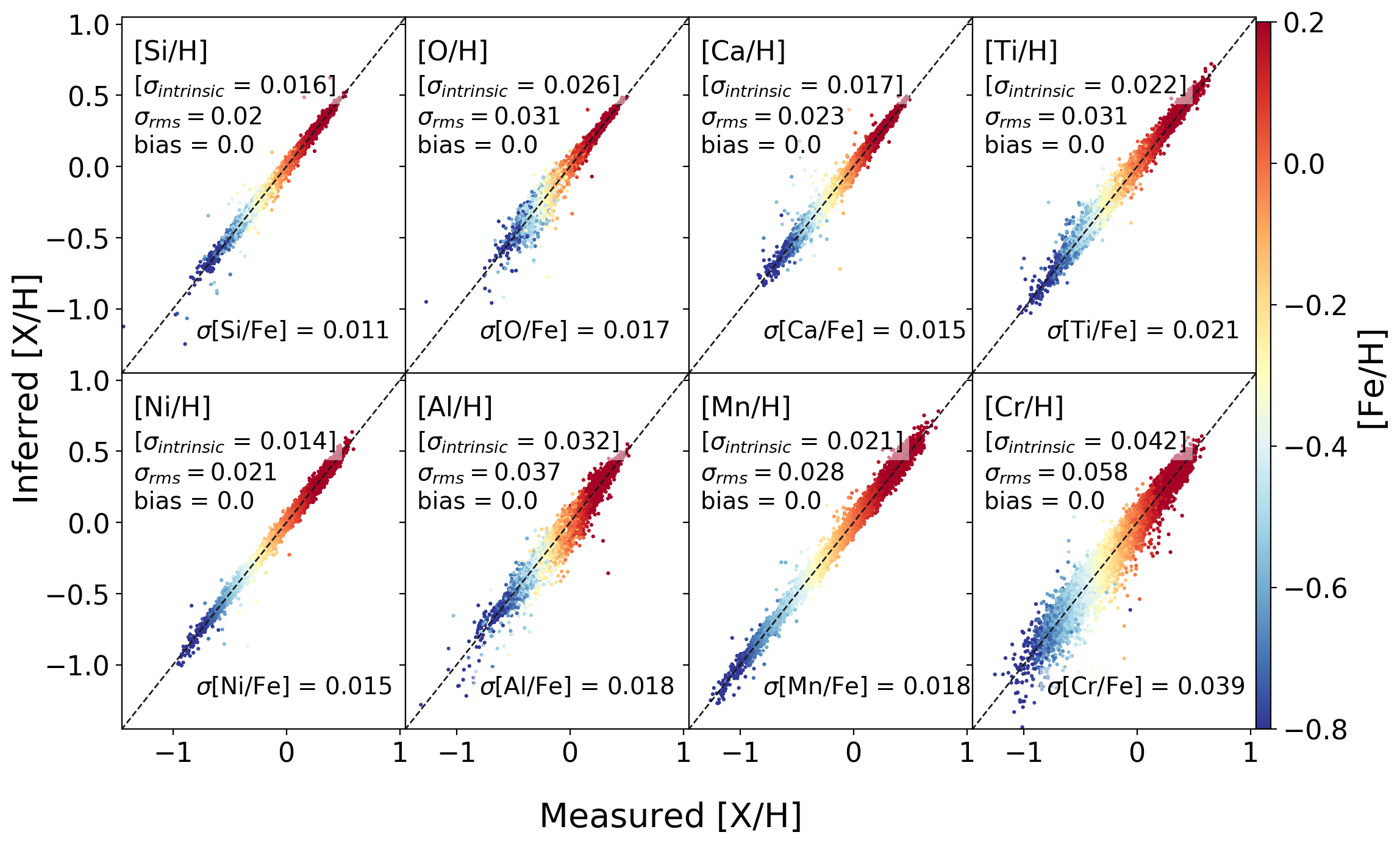}
\caption{The ASPCAP measured abundance versus the predicted (inferred) abundance for each of the eight elements for $\mbox{N}=27,000$ stars in APOGEE. The intrinsic scatter of the prediction is indicated, as is the rms difference between reference (measured) and predicted abundance, and bias. The stars are coloured by [Fe/H] to demonstrate that there is no metallicity bias. The mean error on the measurements is indicated at the bottom right of each sub-panel. Elements are predicted with a precision of between $0.01-0.04$~dex.}
\label{fig:one}
\end{figure*}

\noindent 1. Standardization of parameters: we first standardize the parameters that we wish to use for calculating a distance between each star and every other star. We use these distance measurements to define a neighbourhood for every star. The parameters we select are evolutionary state and the elements from two fiducial supernovae channels: $\vec{Y}$ = (\teff, \logg, [Mg/H], [Fe/H]). To standardize each parameter, we subtract the mean and divide by the standard deviation, calculated for the N stars, for each; 
{{y} = ($Y$ - $\overline{Y})/ \sigma_{Y}$.} \\
2. Distance metric: For each star $n$ in our set of N objects, we determine, via a $\chi^2$ comparison, the distance to each other star in the sample using the four standardized parameters, $\vec{{y}}$ = ( \teff, \logg, [Fe/H] [Mg/H]). For any two stars, the $\chi^2$ value is given by the following equation: 

\begin{equation}
 \chi_{nn'}^2 = \sum\limits_{i=1}^{4} \frac{[y_{ni} - y_{n'i}]^2}{\sigma_{ni}^2 + \sigma_{n'i}^2},
\end{equation}
where the indices $n$ and $n'$ denote the two stars, $i$ is index of each the standardized parameters, and $y_{ni}$ the standardized parameter measurements 
with standardized uncertainty $\sigma_{ni}$.\\ 
3. Define the local model: For each star, $n$, we then take the $k$ nearest neighbours with the smallest $\chi^2$ values (where $k$ sets the size of the neighbourhood). In practice the model is fairly insensitive to the choice of $k$ and between $k = 50-300$ produces comparable results. We select k=100. \\
4. Regression: using the local model for each star of $k$ nearest neighbours, and excluding the star used to define the neighborhood, we use the linear regression function in python's $sklearn$ package \citep{scikit-learn}.   This modeling step learns the relationship between the  parameters of $\vec{{Y}}$ = (\teff, \logg, [Fe/H], [Mg/Fe]) and each of the eight abundances [X/H] (separately), parameterized for each [X/H]) with five coefficients for each local model (the intercept and one for each Y). Then  abundances [X/H] are predicted from the five parameters.\\
5. Predict: for each star, we then (separately) predict each of the eight abundances defined above [X/H] using the four  parameters of $\vec{Y}$ = (\teff, \logg, [Fe/H], [Mg/H]) and the coefficients determined in step (4). This gives our prediction for [X/H], that we can compare to the measured abundance from APSCAP.  \\
6. Repeat: this procedure for every star 1 to N.\\ \\

\section{Results}

\subsection{Local Linear Model Predictions}

The mean summary of our element predictions compared to their measurements using local linear modeling is shown in Figure \ref{fig:one}, for our $\mbox{N}=27,000$ stars. The x-axis shows the ASPCAP abundances for the stars and the y-axis the prediction using a linear regression of the local neighbourhood (modeled with the predicted object removed). A number of measurements are reported in each sub-panel. The name of each element [X/H] is shown in the top left hand corner of each sub-panel. The intrinsic scatter (in a data analysis sense) of each element's prediction, which is the $\sigma_{{intrinsic}}$ is shown in square brackets: This is the root mean square (rms) difference between the predicted and measured abundance [X/H], and the mean measurement uncertainty (which is typically the same for each star) $\sigma_{\textrm{intrinsic}} = \sqrt{\sigma_{\textrm{rms}}^2 - \sigma_{\textrm{measurement}}^2)}$. This bias of the predicted versus the measured abundance [X/H] is also included at top left  (bias =0 as expected from linear regression).  We note that the predicted abundance are unbiased for [Fe/H] $> -1$. However, this is not the case for the halo which is a far more chemically diverse population at fixed (Fe, Mg). The scatter around the 1:1 line is also relatively metallicity independent. The mean measurement uncertainty is reported in the bottom right hand corner of each sub-panel. We note that the measurement uncertainty on the  predictors in our model (\teff, \logg, \feh, \mg) propagate negligibly to the intrinsic scatter that we measure. We note that the systematic uncertainty (accuracy) of the individual element abundance measurements is much larger than the intrinsic dispersion values. Including \teff\ and \logg\ as predictors is done so as to remove systematic biases imputed by stellar model approximations \citep[e.g.][]{ Jofre2019}. We note that while these will remove systematics in the model from model atmospheres, they will not remove biases from anything that is inherited at the level of the spectra itself (see Figure 6).

\begin{figure*}[]
\centering
\includegraphics[scale=0.5]{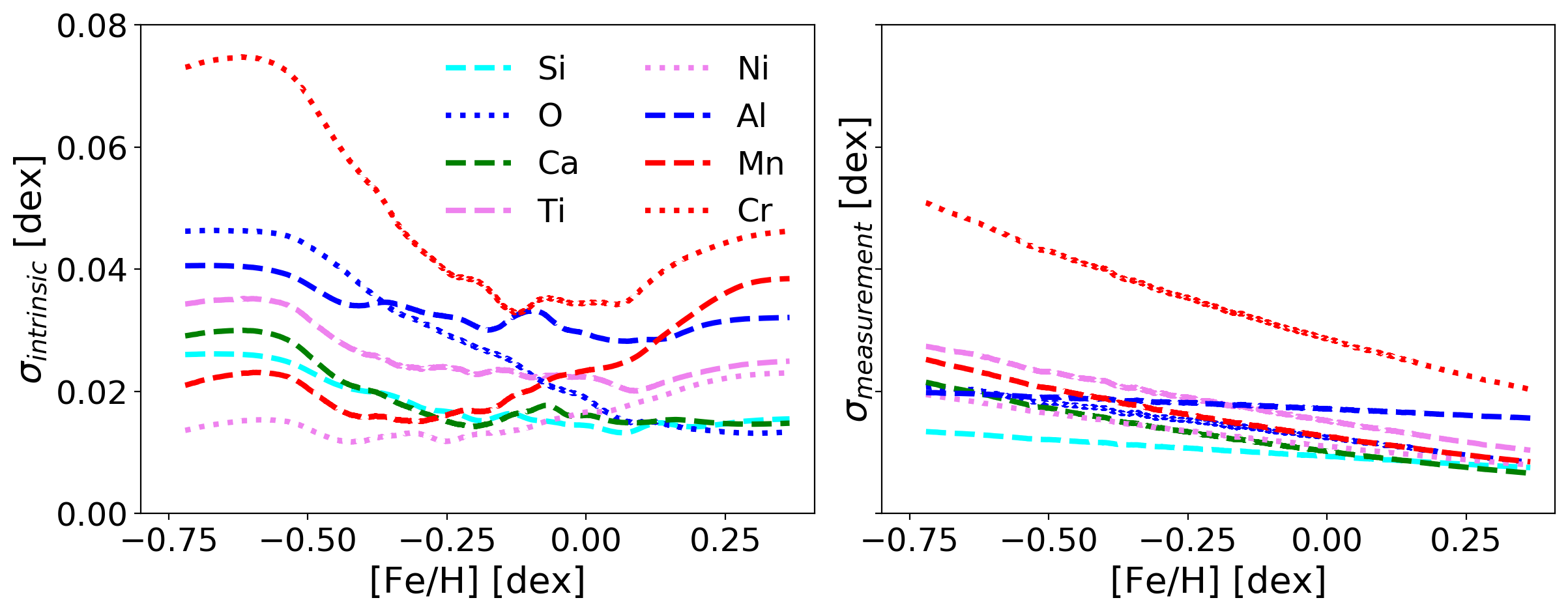}
\caption{At left, the intrinsic dispersion of each element's prediction as a function of [Fe/H] for 27,000 stars with SNR $>$ 200 in Figure \ref{fig:one}. That is, the quadrature difference of the scatter of the measurement compared to the prediction ($\sigma_{\mbox{rms}}$) and the mean error reported on the measurement from ASPCAP ($\sigma_{\mbox{measurement}}$), across [Fe/H]. Different elements show different [Fe/H] relations. The elements Cr and O show an increase with decreasing [Fe/H] and Mn and Ni show an increase with increasing [Fe/H]. Elements of the same colour share source production fractions (see Figure 4). At right, the running mean of the measurement errors ($\sigma_{\mbox{measurement}}$), for each element, across [Fe/H] (the y-axis scale is inherited from the left panel). This demonstrates the element measurement errors increase with decreasing [Fe/H].} 
\label{fig:one_summary}
\end{figure*}

We examine the correlation with intrinsic dispersion as a function of [Fe/H] as shown in Figure \ref{fig:one_summary}. The 1-$\sigma$ intrinsic dispersion of the [X/H] measurement compared to the prediction is shown for each element. There are unique trends in each element; some on the order of a few percent and some on the order of a factor of two, across [Fe/H]. The elements Cr and O show the most dramatic increase with decreasing [Fe/H] and the element Mn shows the most significant increase at higher [Fe/H]. Beyond noting that the intrinsic dispersion varies across metallicity [Fe/H], we do not draw strong astrophysical conclusions from these trends. This is because the total scatter of the [X/H] prediction and subsequently the intrinsic dispersion values that we calculate are sensitive to the accuracy of the abundance measurement errors, which vary across the [Fe/H] plane. For all elements, the measurement errors increase with decreasing metallicity, with the mean error for elements Cr and O increasing the most dramatically by a factor of more than two.

\begin{figure*}[]
\centering
\includegraphics[scale=0.7]{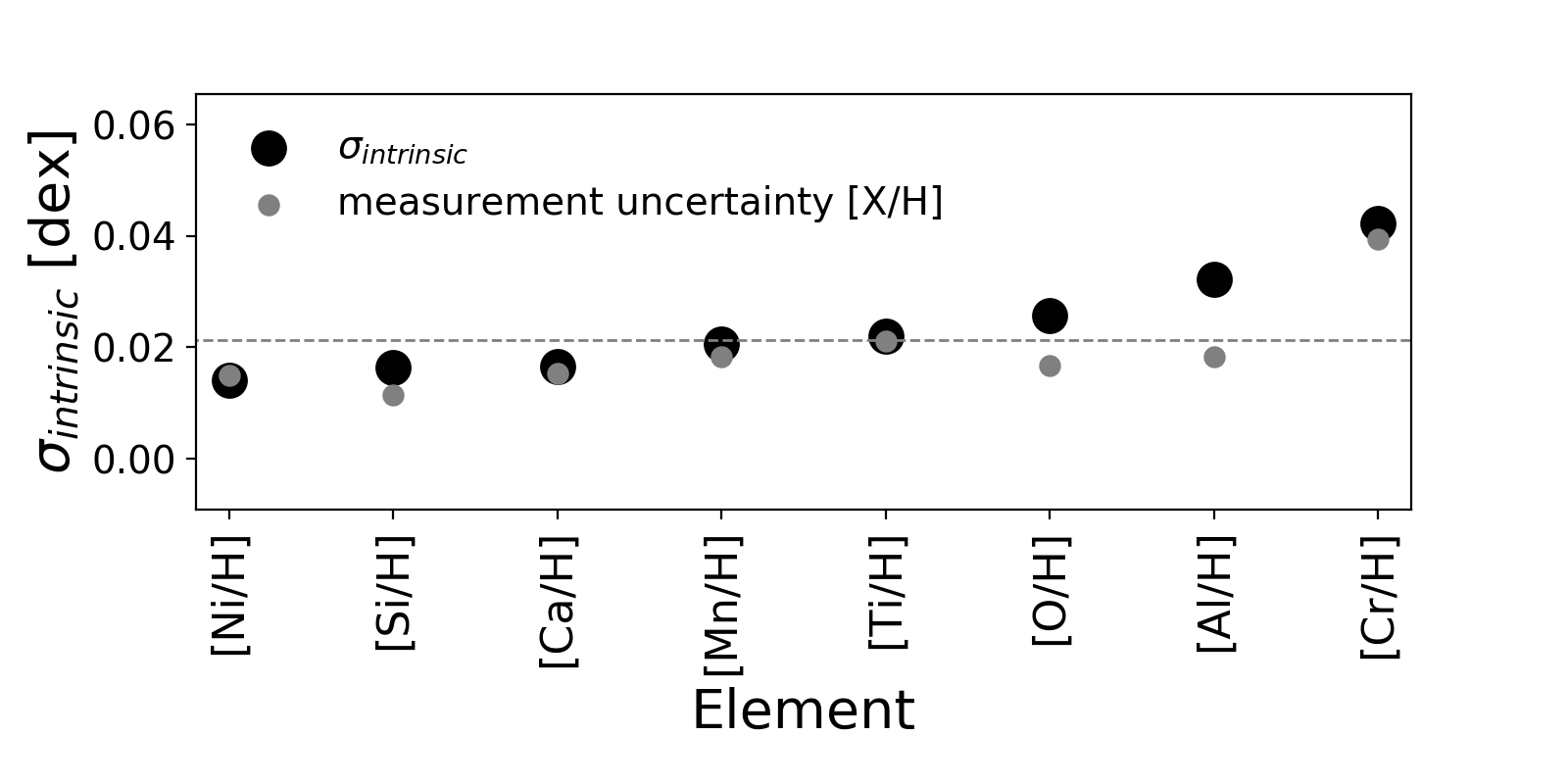}
\caption{The summary of the mean intrinsic dispersion of the prediction compared to the measurement shown in Figure \ref{fig:one}, for each element [X/H]. The dashed line represents the median and the small grey points show the measurement uncertainty.} 
\label{fig:two}
\end{figure*}

Each element's mean intrinsic dispersion across all [Fe/H] (or intrinsic scatter) from Figure \ref{fig:one} are compared in Figure \ref{fig:two}. The dashed line indicates the median intrinsic dispersion for these eight elements,  which is $\sigma_{intrinsic} \approx 0.021$~dex. That is, using (Fe, Mg) alone the other elements produced in supernovae can be predicted on average to this precision, but with a range of $0.014$~dex (for Ni) to $0.04$~dex (for Cr).

 The intrinsic scatter for each element [X/H] is a quantification of the effective individuality of the element. That is, how much additional information it captures beyond (Fe, Mg) alone. The intrinsic scatter also represents the diversity of the star forming environment. For example, given the (Fe, Mg) of a star alone, to learn something additional from Si, from Figure \ref{fig:one}, we see that we need to measure the Si abundance to $< 0.015$~dex.

\subsection{Residuals: what do these mean?} 

Elements are individuals \citep[e.g.][]{Freeman2002, Buder2018, CK2020, Blancato2019, Ting2021}, but as summarised in Figure \ref{fig:two}, only marginally (compared to typical measurement uncertainties), for supernovae sources. While the ten elements considered, of Fe, Mg, Si, O, Ca, Ti, Ni, Al, Mn, Cr,  are produced in supernovae, this is (i) via different SNII compared to SNIa fractions, (ii) with different [Fe/H] dependencies and (iii) with different mass dependencies \citep[e.g.][and references therein]{Woosley1995, CK2020, Matt2021}. For each element shown in Figure \ref{fig:one}, we calculate the residual of that element =  measured abundance from ASPCAP $-$  model's prediction. A positive residual means the measured abundance is higher than predicted by the local neighbourhood and a negative residual means the abundance is less than that expected in the local neighbourhood. As the residuals quantify how far each star deviates from the model's prediction, in principle these may be valuable tracers of birth environment, that are independent of (Fe, Mg). That is, because the element abundances ([Fe/H], [Mg/H]) are the predictors, the metallicity dependence of the yield has been removed, so scatter is caused  by other conditions of the star forming environment. 

\begin{figure}[]
\centering
\includegraphics[scale=0.5]{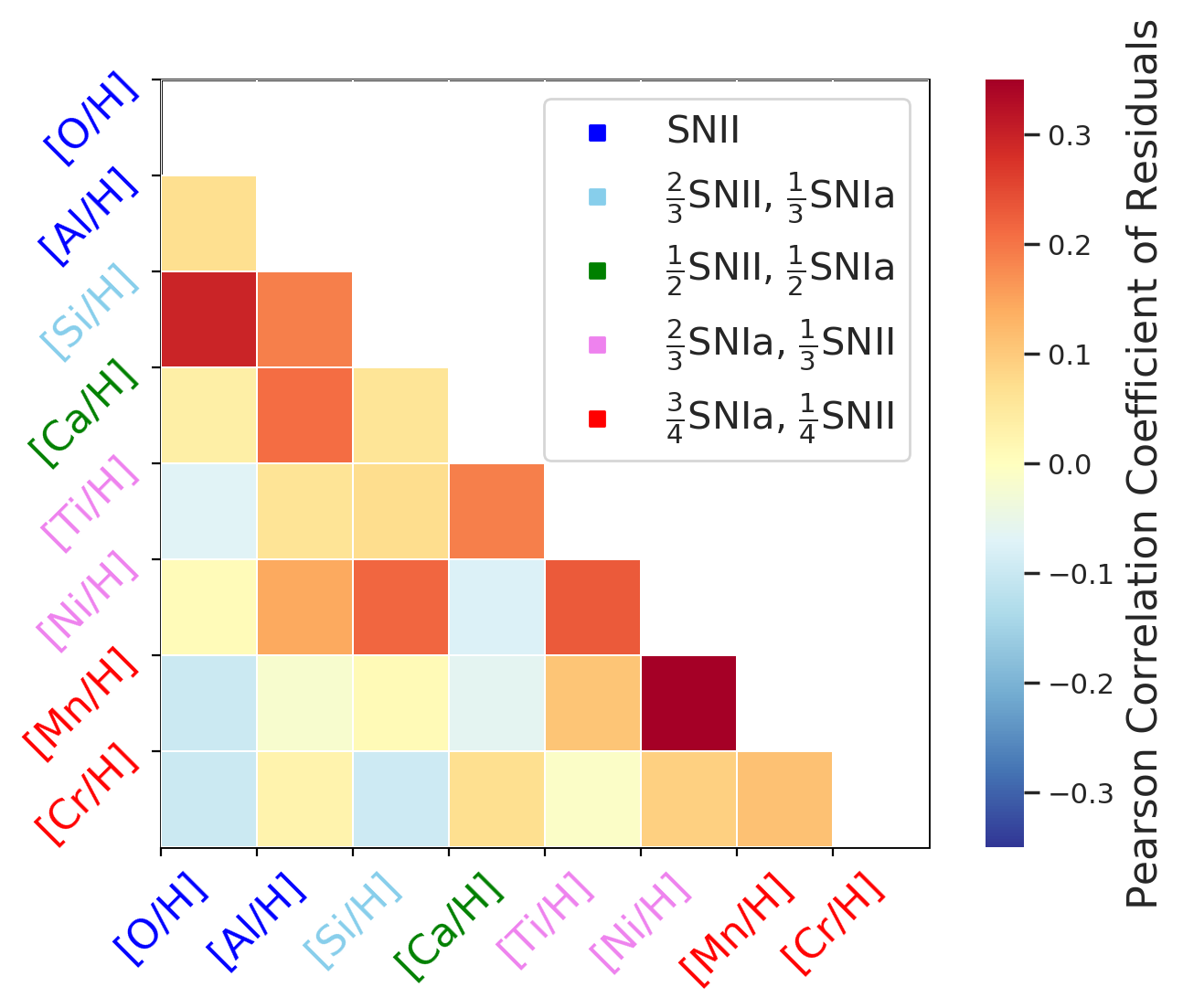}
\caption{The correlation matrix for the eight element residuals determined from the models' prediction compared to the [X/H] measurement as summarised in Figure 1.  The residual pertaining to each element is for the element indicated on the axes. The element source contribution is indicated by the color of the label, shown in the legend \citep[][]{JJ}.} \label{fig:cors}
\end{figure}

Figure \ref{fig:cors} shows a matrix of the correlations of the residuals of the predictions.  This Figure shows that intra-family elements  (e.g. Si and O, produced predominatly in SNII) appear to be correlated and inter-family (e.g. Cr and Si, produced predominantly in SNIa and SNII, respectively) appear anti-correlated. Note, all correlations are fairly weak (a Pearson correlation coefficient of 1 would be a perfectly correlated set of two error-free distributions). 

 We now investigate correlations of the residuals with age and present-day orbital properties (time and place). The correlations between the element residuals, and their trends with age and orbit are revealing as to the difference in the star forming environment that has given rise to their abundances. Furthermore such correlations of the residuals would indicate additional resolving power in the eight elements in Figure \ref{fig:one} to identify birth environment beyond Fe and Mg alone \citep[][]{Ting2021}.

 \subsection{Correlations of residuals with age and orbit} 

\begin{figure*}[]
\centering
\includegraphics[scale=0.31]{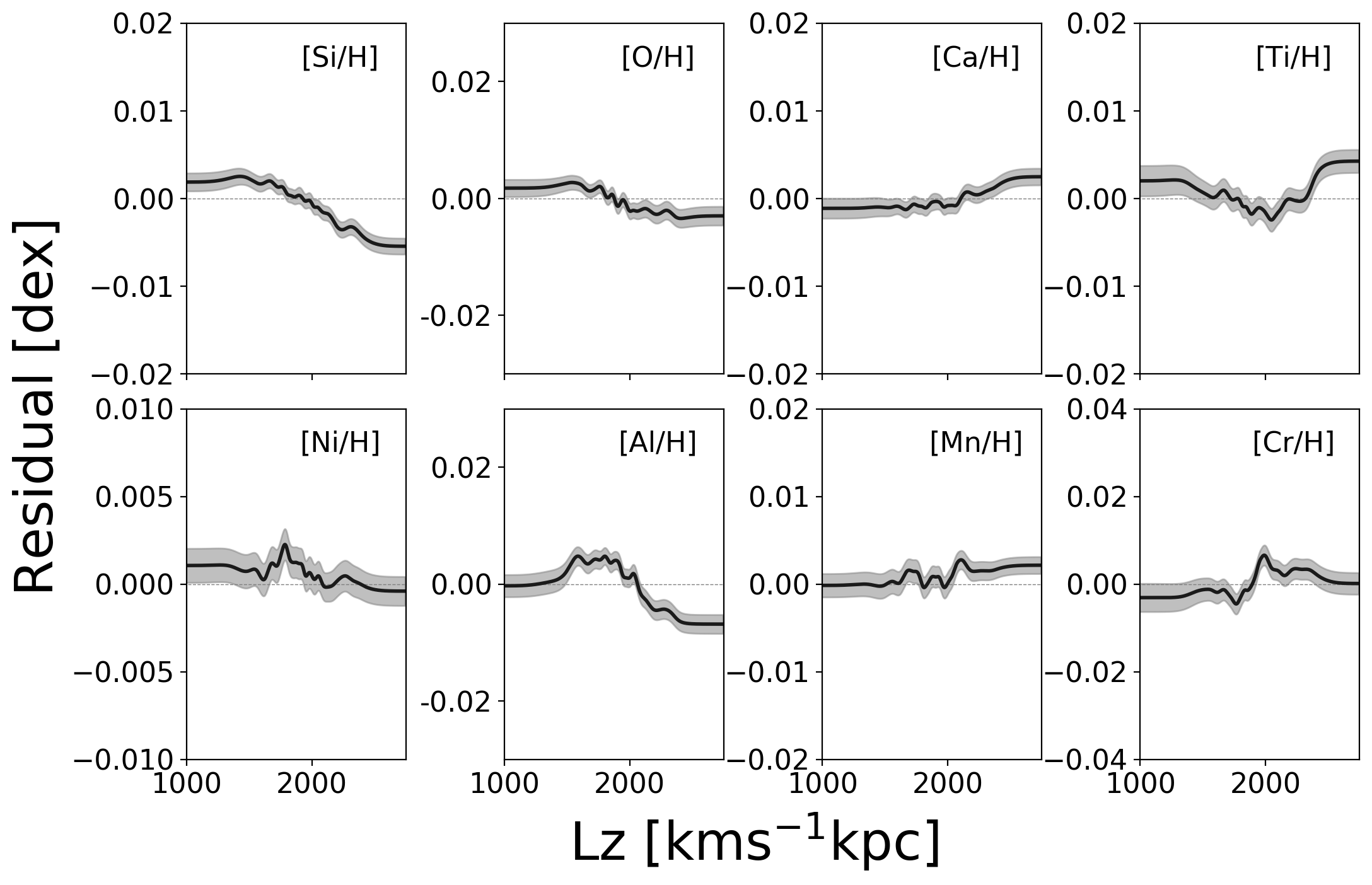}
\includegraphics[scale=0.31]{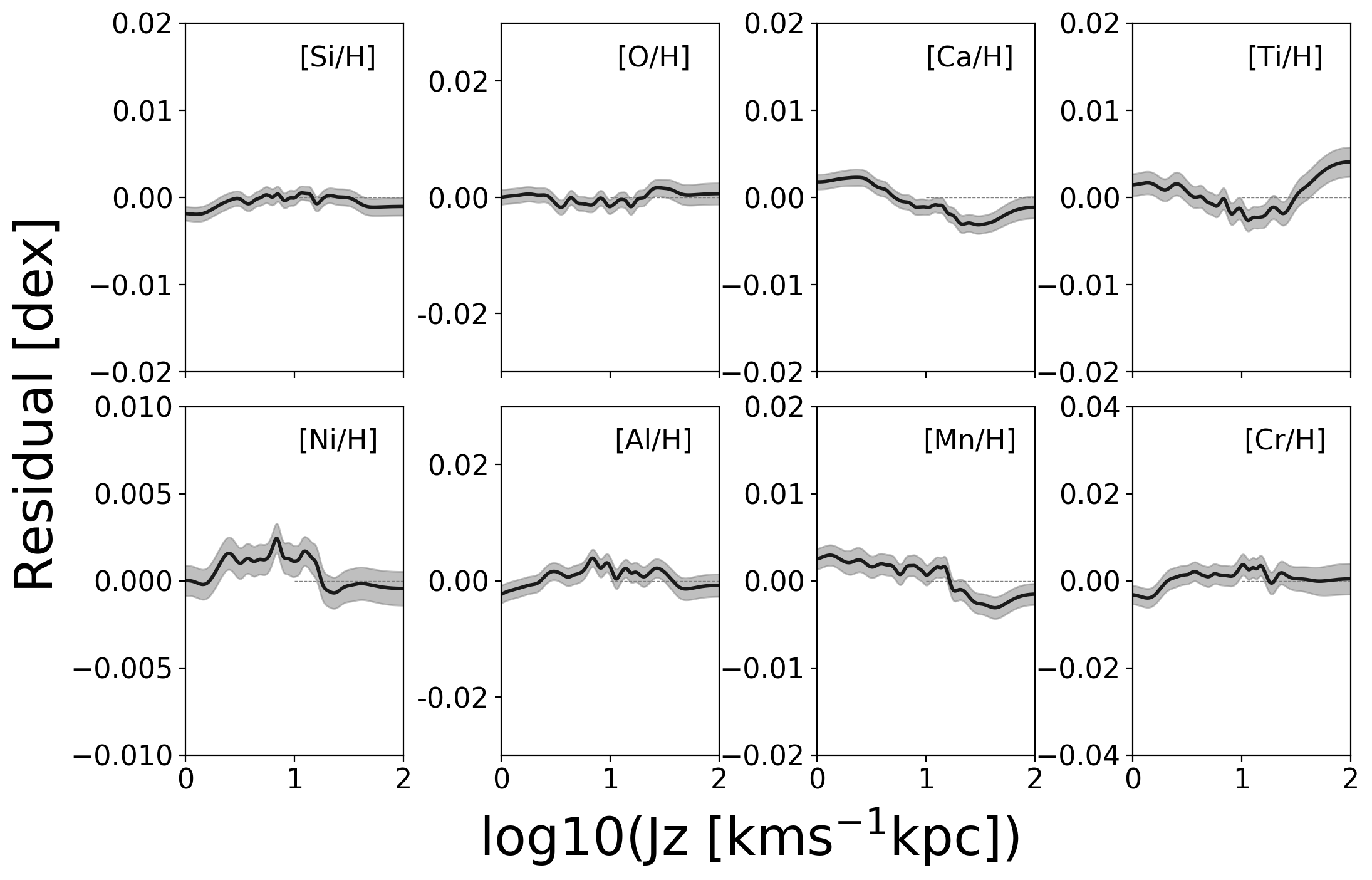}
\caption{The running mean of the angular momentum $L_z$ at left, and the log of the vertical action $J_z$ at right, versus the residual amplitude for each element, with the y-axis for each scaled to the intrinsic dispersion, showing that the astrophysical correlation with $L_z$ represents only a fraction of the residual amplitude. The shaded region represents the confidence on the mean. (Note only 10 percent of stars have $J_z > 30~\kmskpc$)} 
\label{fig:three}
\end{figure*}

Figure \ref{fig:three} shows the correlation between the element abundance residuals and the angular momentum (at left) and vertical action (at right), $L_z$ and $J_z$, respectively. We do not include the radial action $J_r$, as the residual trends are flat and the range is limited, given as these are disk stars.  The (smoothed) running mean of action-residual trends are shown and the shaded regions represent the 1-$\sigma$ confidence on the mean. The axes are scaled to be the amplitude of the intrinsic dispersion measured as reported in Figure \ref{fig:one}. The axis scaling  highlights that the spatial/orbital correlations explain only (small) part of the total magnitude of this scatter. 

\begin{figure*}[]
\centering
\includegraphics[scale=0.31]{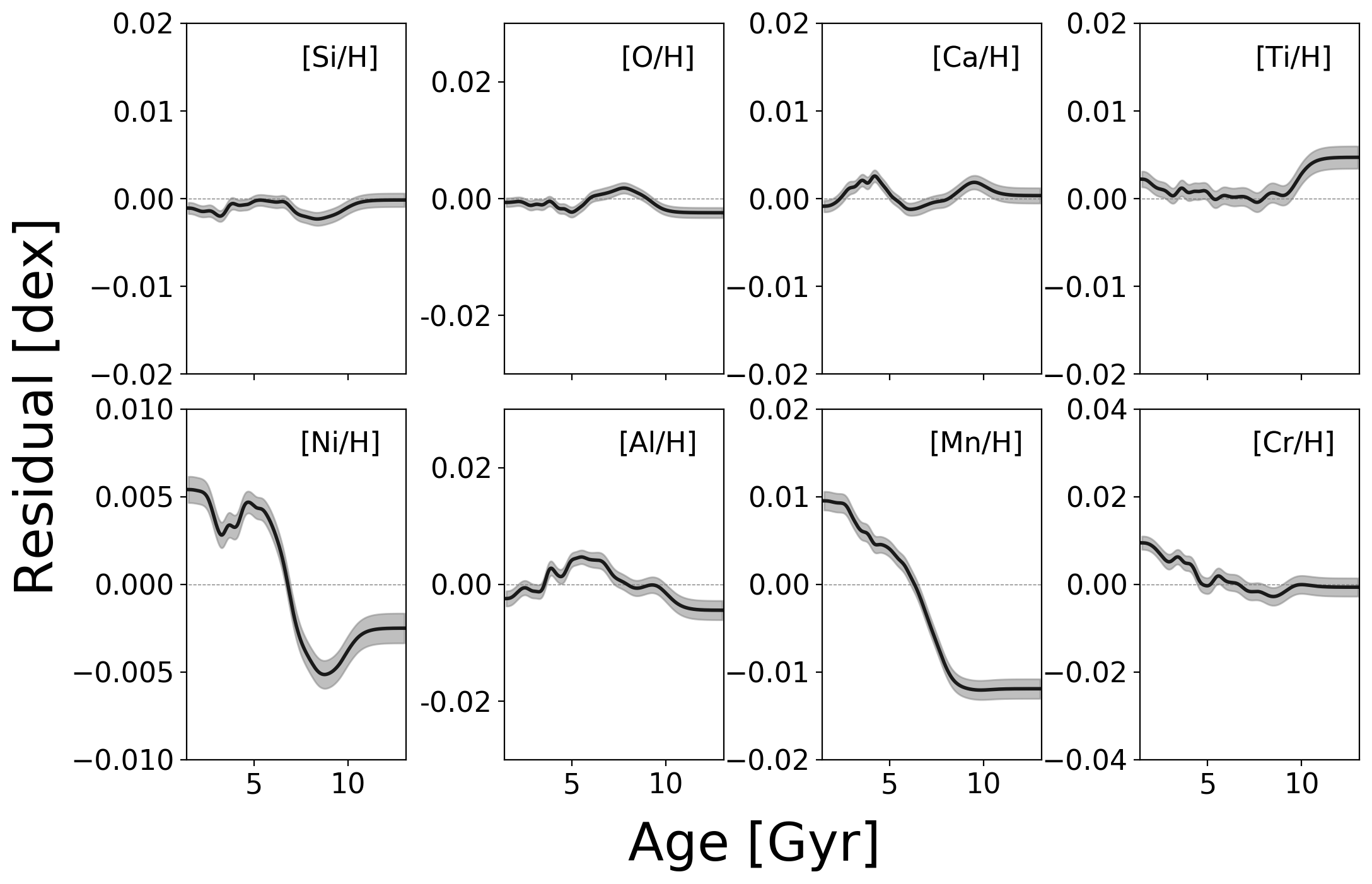}
\includegraphics[scale=0.31]{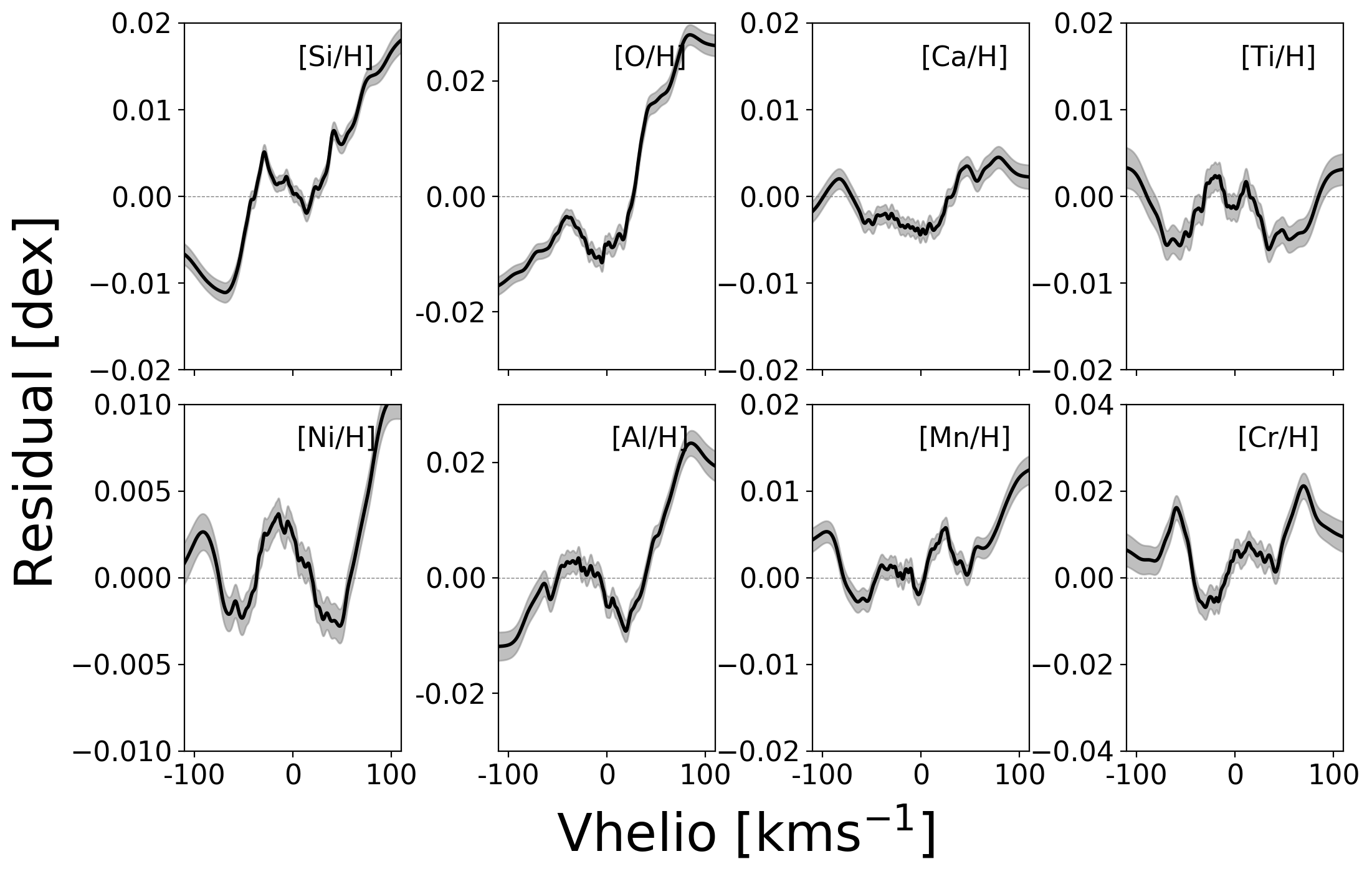}
\caption{At left, the running mean of the age versus the residual amplitude for each element with the y-axis for each scaled to the intrinsic dispersion amplitude. There are $\sim$15,000 stars used in this figure. At right, the velocity correlation for a sample of $\sim$80,000 stars with SNR $>$ 100. This figure at right demonstrates why we implement a velocity cut to calculate the amplitude of our intrinsic dispersion for these elements. The shaded region represents the confidence on the mean.} 
\label{fig:four}
\end{figure*}

Similarly to Figure \ref{fig:three}, Figure  \ref{fig:four}, at left, shows the residual of each element versus age.   There are $\sim$15,000 stars with ages \citep{Lu2021} that are used to make the panel showing the age correlation. The right panel of Figure \ref{fig:four} shows a larger sample of 80,000 stars with an SNR cut of $>$ 100 (and no velocity restriction imposed). Figure \ref{fig:four} at right shows the correlation between residual amplitude for these 80,000 stars and heliocentric velocity. This is the most significant variable accounting for correlations in the abundance residuals -- and likely due to nuisance signals in the spectra near the limits of the precision. Initial results suggest that residual telluric and sky line and other non-stellar features remain in the \apogee\ spectra at the level of $1-2$ percent across many wavelengths. Because these features are in the (approximate) observer frame, the stellar rest frame spectra have contamination (which impacts measured abundances) that correlates with stellar heliocentric velocity (see McKinnon et al., in preparation). The relations are non-monotonic and are non trivial to remove, and therefore separate from the abundance measurements themselves. 

We note that if we repeat the exercise of the calculating the residuals of the [X/H] predictions and subsequently examining orbit and age correlations with a lower fidelity sample e.g. SNR $>$ 100 and remove any velocity limit, we find effectively the same results to those shown for our higher fidelity sample here. However, we measure higher intrinsic scatters for a few of the elements (shown in the next Section in Figure \ref{fig:clusters}).

In summary, these correlation Figures \ref{fig:three} and \ref{fig:four} demonstrate that the astrophysical origins of the intrinsic scatter of each element [X/H] around the model's prediction from (Fe, Mg) can only partially be attributed to birth time and present orbit/location.

\subsection{Intrinsic variance not explained by astrophysical properties}

 We see from Figures \ref{fig:three} and \ref{fig:four} that orbital parameters and age correlations explain only part of the intrinsic scatter around the abundance predictions shown in Figure \ref{fig:one} (residual amplitude). To test how much the intrinsic scatter decreases once we remove known correlations, we add to the regression model the additional parameters shown in Figures \ref{fig:three} and \ref{fig:four}, of $J_z$, $L_z$, age and $\vhelio$. We do this to try to assess how much of the variance remains once we account for other sources of correlation. We notice that additional small correlations are seen in the residuals with mean fiber number and SNR, and subsequently add these as additional parameters to our model as well. Furthermore, as we are interesting in finding a model with a minimum level of  intrinsic dispersion of the measurement minus the predicted [X/H], we explore additional parameter cuts. We find that introducing a restricted vertical extent of the stellar orbits, $J_z$, gives another $\sim$ 5 percent reduction in $\sigma_{intrinsic}$. We therefore restrict this sample to disk stars with $J_z < 30~\kmskpc$, which leaves a total of $\sim$13,000 stars.

\begin{figure*}[]
\centering
\includegraphics[scale=0.75]{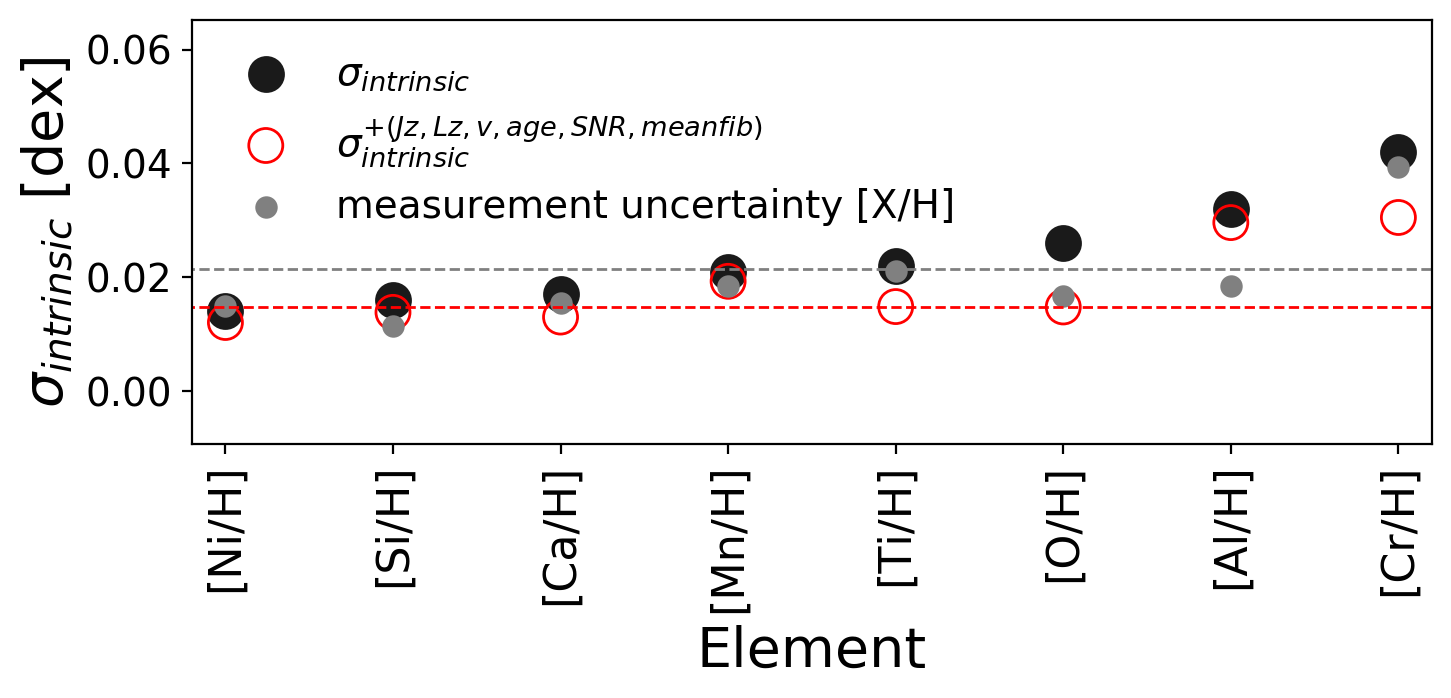}
\caption{The intrinsic dispersion of the prediction compared to the measurement as shown in Figure \ref{fig:one}, for each element (black) compared to a model with additional parameters included (red). The dashed grey line represents the mean intrinsic dispersion, using four parameters for the prediction, from Figure \ref{fig:two}. The red line shows the mean intrinsic dispersion with additional parameters included. The small grey points show the measurement uncertainty. Note the model shown in black points is calculated using 27,000 stars and the model in red with the subset of stars for which ages are available of 13,000 stars. Residual trends shown in Figures 5 and 6 are removed with the model in red, with the exception of partial age trends remaining for Ni and Mn.} 
\label{fig:endfig}
\end{figure*}

 In Figure \ref{fig:endfig}, we show the summary of the intrinsic dispersion for each element from the model with the additional parameters included as predictors. Including this additional set of parameters in the model reduces the intrinsic dispersion of the eight elements to a median level of $\approx 0.015$~dex. This demonstrates that an estimate of the residual variance that is not able to be explained with age or orbit up to $\approx 0.015$~dex for these eight elements. The local linear model does not, however, entirely remove the correlations of these variables. Specifically, for Ni and Mn partial trends with age persist at the $< 0.005$~dex level. Therefore, the median intrinsic dispersion of $\approx 0.015$~dex represents an upper limit.  We note that for this model we found a larger neighborhood, with $k$ = 300, gave a marginally lower intrinsic dispersion for the element Cr in particular, by about 15 percent. This is not surprising that a larger $k$ might be favoured given the larger number of variables used in the model.
 
 We verify that the smaller intrinsic dispersion we obtain is a consequence of including additional predictors in the model by using our initial four parameter model applied to this subset of 13,000 stars. The intrinsic dispersion obtained with the smaller subset of 13,000 stars and four parameter model is near-identical to that obtained with the 27,000 fiducial sample, with a median $\sigma_{\mbox{intrinsic}} = 0.20$~dex compared to $\sigma_{\mbox{intrinsic}} = 0.21$~dex for the full sample. The subsequent reduction in the scatter of the elements around the model with additional parameters, is therefore a consequence of including these additional correlated variables in the model regression and not the different samples.

 To validate these results, we do an additional calculation as follows, to estimate how much of the overall intrinsic scatter we measure for each element is not a consequence of correlation with physical parameters and velocity: $L_z$, $J_z$, age and $\vhelio$. For each of these four parameters, we subtract in quadrature the average 1-$\sigma$ dispersion around the mean-trends reported in Figures \ref{fig:three} and \ref{fig:four}, from the rms-dispersion as measured around the prediction of the model in Figure \ref{fig:one}, for each element. This gives an overall measure of the  amplitude of the scatter that is explained by each of $L_z$, $J_z$, age and $V_{\mbox{helio}}$ (combined with the measurement errors) for each element. The quadrature sum of these for each element represents the total scatter (combined with the measurement errors) accounted for by these joint parameters (ignoring any correlation between them, so as to obtain a lower limit), $\sigma_{\mbox{corr}}$. We then take the quadrature difference for each element between the intrinsic dispersion of the model's prediction summarised in Figure \ref{fig:two} and the quadrature sum of the scatter explained by the four parameters;  $\sqrt{\sigma_{\textrm{intrinsic}}^2 - \sigma_{\textrm{corr}}^2}$.
 This gives us how much intrinsic variance is not explained by these parameters. On average, we find  this to be $\approx 0.01$~dex, for the eight elements. 
 
 Is it possible that there are correlations with birth radius that would explain the remaining intrinsic scatter, but as stars migrate over time this is simply lost \citep[e.g.][]{Selwood2002, Roskar2008}? We examine the $L_z$ and $J_z$ trends for only the youngest populations. Although the residual-action gradients are steeper (see Figure \ref{fig:three_appendix} in the Appendix), these are still only are able to explain a fraction of the intrinsic scatter measured in each element. We therefore think this is unlikely, as this would be preserved in the age-residual gradients, particularly for the youngest stars. Stars of a given birth radius likely follow tracks in [Fe/H]-age \citep[][]{Minchev2018, Frankel2018}. Yet, we see from Figure \ref{fig:four} that the age correlations are rather weak (including in narrow ranges of [Fe/H]), with the exception of Mn and Al which still only account for part of the variance. We therefore conclude that $\approx 0.01-0.015$~dex is an effective measure of the amplitude of dispersion in these eight elements in the disk at birth time and radius.

\subsection{The residual scatter of open clusters}

We test the intrinsic scatter of residual abundances for the eight individual elements, [X/H], using known sites of common birth origin -- open clusters. Cluster members are selected from our set of $\sim$27,000 stars as those with a membership probability of $> 0.7$ as reported in the APOGEE OCCAM open cluster catalogue \citep[][]{occam}. This gives us a total of 46 stars distributed among the following four clusters: NGC 188, NGC 2682, NGC 6705 and NGC 6819. For each element, in each cluster, we calculate the mean measurement error for the set of stars, and the 1-$\sigma$ dispersion of the residuals in each [X/H]. The quadrature difference of the scatter of the residuals and the measurement uncertainties gives an average intrinsic dispersion of $\approx 0.01$~dex for the eight elements and four clusters (with a range range of $0.005$~dex for Si to $0.02$~dex for O). This mean value of $\approx 0.01$~dex is consistent with our inference from the field stars, of $\approx 0.01-0.015$~dex scatter in elements at fixed birth radius and age. 

\begin{figure*}[]
\centering
\includegraphics[scale=0.5]{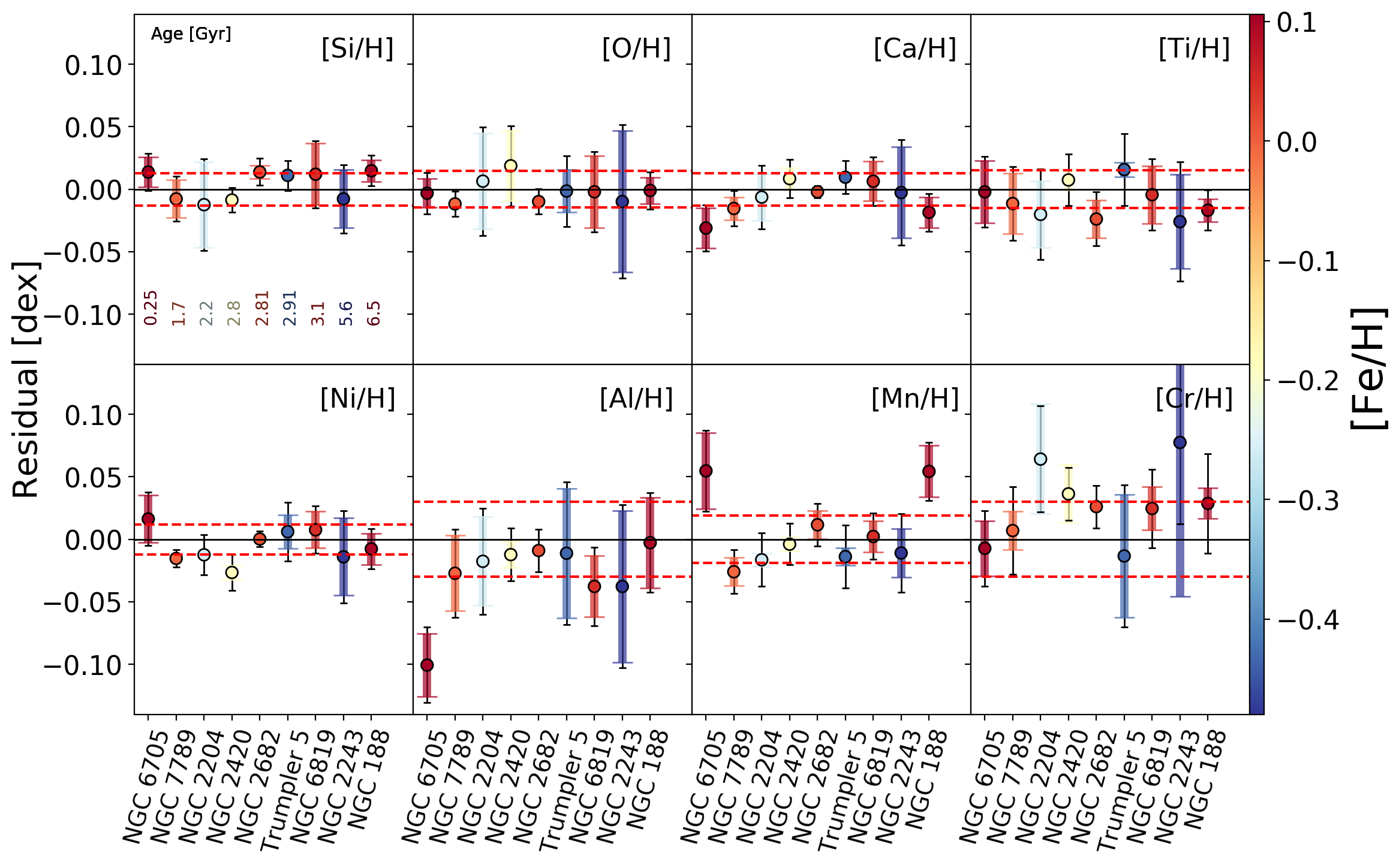}
\caption{The distribution of the residuals (of the measurement $-$ prediction) for each element [X/H] in the nine clusters (each with eight or more members). These are ordered from youngest to oldest cluster (as indicated in vertical annotation, see text). The points show the mean of the residual distribution, colored by [Fe/H]. The thin error-bars show the 1$-\sigma$ dispersion around the mean. The wider colored error-bars show the intrinsic dispersion of the residual distribution (from the quadrature difference of the dispersion of the 1$-\sigma$ residual distribution and the mean error of the cluster stars, for each [X/H]). The red dashed lines indicate the intrinsic dispersion from the field model for 13,000 stars shown in Figure 7. The clusters have a range of intrinsic dispersion values, and overall are not dramatically dissimilar to the field comparison shown in the dashed red lines. The median intrinsic dispersion of the clusters for these eight elements is $\approx 0.015$~dex (correspondingly, $\sim$ 50 percent of the cluster intrinsic dispersion measurements are smaller than the field stars and $\sim$ 50 percent are larger). Note the bias around the zero-line in the mean residual amplitude for many elements in all clusters. This is in part explained by the residual trends documented in Figures \ref{fig:three} and \ref{fig:four}, with exceptions.} 
\label{fig:clusters}
\end{figure*}

We now use a more lenient selection of stars to study the residuals in more open clusters. We use the larger set of 80,000 stars with $SNR$ $>$ 100. From this sample, we obtain 107 stars from nine different clusters with a membership probability of $> 0.7$ and more than eight stars per cluster. Figure \ref{fig:clusters} shows a summary of the residual amplitude away from the model's prediction, for each element, in each cluster. These are ordered from youngest to oldest cluster \footnote{see \citep{Salaris2004, JP1994, Trumpler5, NGC188, NGC2682, NGC6819}}. For reference, in the context of Figures \ref{fig:three} and \ref{fig:four}, these clusters have the following distributions in parameters: 
Age = $3 \pm 1.8$~Gyr~($0.25-6.5$~Gyr), $J_z = 10 \pm 9~\kmskpc~(0.26 - 26~\kmskpc$), $L_z = 2135 \pm 390~\kmskpc~(1468 - 2812~\kmskpc$), $\vhelio = 29\pm 48~\kms~(-54 - 94~\kms$). Although selected with a minimum SNR $>$ 100, 70 percent of the open cluster stars have SNR $>$ 200.

\begin{figure*}[]
\centering
\includegraphics[scale=0.5]{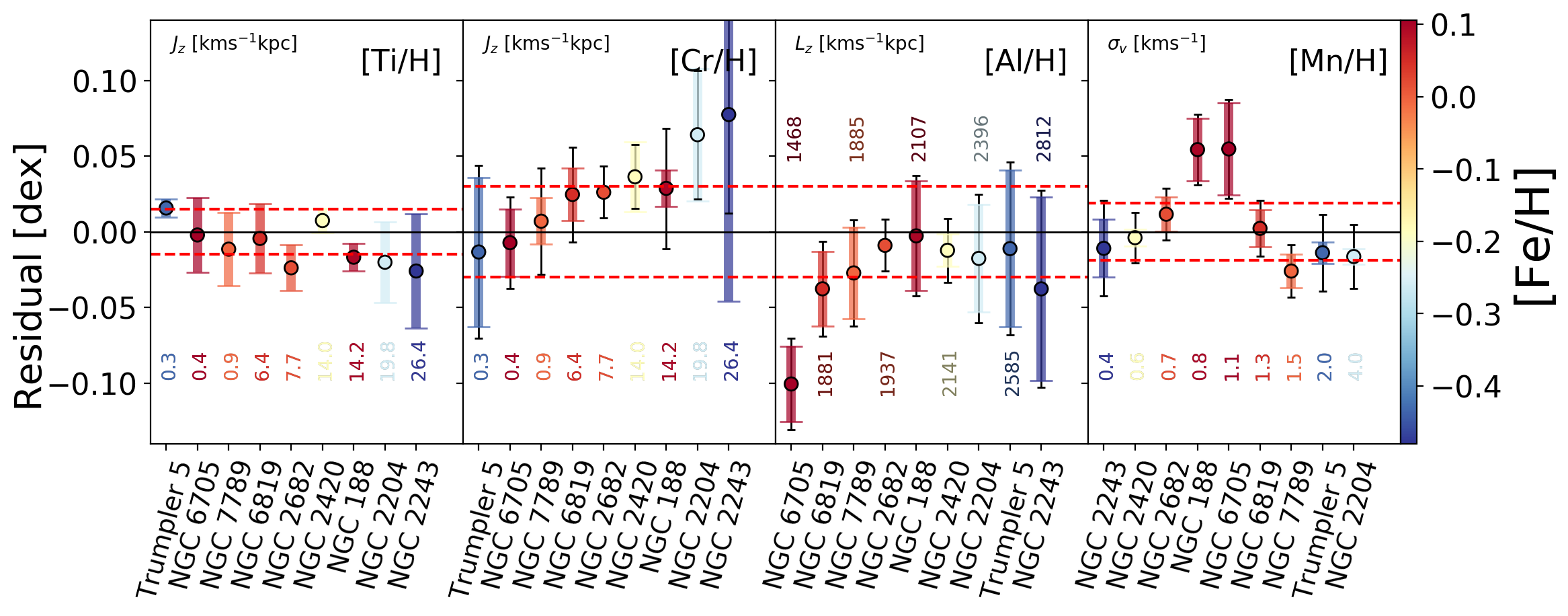}
\caption{Four elements selected from Figure \ref{fig:clusters}, sorted by physical cluster parameters as indicated in the top left hand corner of each panel. The corresponding amplitude of these parameters is reported with vertical text. The two left panels, which report the Ti and Cr residuals, are sorted by $J_z$. The next panel, at right, showing the Al residual, is sorted by $L_z$. At right, the residual of Mn is sorted by cluster heliocentric velocity dispersion; the most-metal-rich clusters have the most similar velocity dispersion values, and anomalously high residuals in Mn. } 
\label{fig:clusters_sorted}
\end{figure*}

In Figure \ref{fig:clusters}, the mean residuals are colored by mean [Fe/H] of the cluster stars (which range from $-0.5$~dex $<$ [Fe/H] $<$ 0.1~dex). The thin error bars around the mean report the 1-$\sigma$ dispersion of the residual distributions. The thick, coloured error bars show the intrinsic dispersion of the residuals (from the quadrature difference of the 1-$\sigma$ dispersion and the mean error of the cluster stars). The red dashed lines in Figure \ref{fig:clusters} show the intrinsic dispersion measurements calculated for each element using the set of stars with the model from Figure 7. As described in Section 4.3, this comprises a sample of 13,000 field stars of the disk (SNR $>$ 200), with age, orbital actions (and also velocity) included in the model for as predictors for each element. This serves as a baseline comparison for the open cluster sample results  (where the stars in open clusters have the same orbital and age properties so these are implicitly accounted for in examining the clusters). About 50 percent of the time, the intrinsic dispersion of the open cluster residuals exceeds that of the field model. The mean intrinsic dispersion of the residuals for the eight elements and nine clusters is $\approx 0.015$~dex (with a variation of $\pm 0.02$~dex on average for the distribution of all clusters and all elements, with individual elements measuring between $0-0.1$~dex). The smaller variance compared to the field (red dashed lines in Figure \ref{fig:clusters}) for some clusters in some elements indicates that the cluster comprises a more homogeneous environment, where elements are more precisely predicted by (Fe, Mg),  compared to the field. However, on average, the  intrinsic dispersion of the clusters is not markedly different from the field. 

Looking at the distribution of the mean residual values in Figure \ref{fig:clusters}, it is clear that the cluster stars have biased mean residuals with respect to the field.  On average, the intrinsic dispersion of each measurement is smaller when the bias is larger.  The mean of the residuals in the field sample is shown with the black line at zero, for reference. For individual elements, the set of clusters are approximately distributed around zero-mean, with the exception of Al and Cr.  Individual clusters themselves, however, show bias away from zero-mean. This must be due to some difference between the clusters and the field distribution. Part of the bias in each element for the clusters can be explained by the correlations reported in Figures \ref{fig:three} and \ref{fig:four}, and the interplay between these correlations. 

Figure \ref{fig:clusters_sorted} shows the residuals of the four elements with the largest scatter in the cluster means (Ti, Al, Mn and Cr) ordered in increasing vertical action, $J_z$ (for Ti and Cr), angular momentum, $L_z$ (for Al), as well as dispersion in heliocentric velocity, $\sigma_{V}$ (for Mn). We look at the correlation with velocity dispersion due to its association with tracing cluster mass \citep[e.g.][]{Poov2020}. Figure \ref{fig:clusters_sorted} reveals that the correlations with the orbital actions appear stronger than in the field. The residuals for Al show a relationship across $J_z$ that echoes some of the structured variation in the mean seen in the young field stars of the disk, in Figure \ref{fig:three_appendix}. However, NGC 6705 has an anomalously low Al residual and the mean residual value for three of the clusters exceeds the 1-$\sigma$ intrinsic dispersion of the field. The residuals for the elements Ti and, in particular, Cr, show a correlation with $J_z$. The correlation between the residual of Cr and $J_z$, in particular, is not similarly seen in the field sample. The relationship between the residual in Cr and $J_z$ suggests that the amplitude of Cr at fixed (Mg,Fe) is a tag of birth height from the plane of the disk, that is erased with cluster dissolution and heating.  The element residuals for Cr and Al have the largest intrinsic dispersion values, by a factor of about two, compared to the other elements (with $\sigma_{\mbox{intrinsic}} = 0.03$~dex). The correlations with the (presumably birth) orbital actions seen here for these elements likely account for part of that scatter measured in the field.

The residuals of the element Mn show a correlation with age in Figure \ref{fig:clusters} that is overall consistent with the trends seen in Figure \ref{fig:four}, with the exception of NGC 188, which is the oldest cluster with age $= 6.5$~Gyr. From Figure \ref{fig:four}, old stars are biased to have negative residual values for Mn. Yet, the mean Mn residual for NGC 188 is $\approx 0.05$~dex (and near-identical to that of the youngest cluster, NGC 6705). Interestingly, the two clusters, NGC 188 and NGC 6705 are the most metal rich clusters and show similar velocity dispersion values. These two clusters also have anomalously low residuals in Ca. 

We examine the cluster with the highest overall intrinsic dispersion in the elements, NGC 2243. This cluster shows a dramatically larger intrinsic dispersion in the residual of Cr, compared to the other clusters. Interestingly, this cluster has a very high radial action compared to the other clusters, with $J_r = 200 \pm 54~\kmskpc$. The other eight clusters have
$J_r = 5-35~\kmskpc$. NGC 2243 is also the most metal-poor cluster and Figure \ref{fig:one_summary} shows that for the field sample, the residual scatter for six of the eight elements increases with decreasing metallicity, most substantially for the element Cr (by a factor of about two between $\feh = 0$ and $\feh = -0.5$).

In summary, the analysis of the open clusters shows that there is, on average, an intrinsic scatter of  $\approx 0.01-0.015$~dex in the individual element residuals [X/H] predicted from (Fe, Mg) in groups of stars known to be born together. This is the same order of magnitude as the intrinsic dispersion in the field, measured after accounting for orbital and age correlations. However, the clusters also show mean residuals that are biased away from zero. Part of this is explained by the expected correlations between the residuals and age and orbital actions. However, the correlations appear to be more substantial in the open clusters compared to the field, for some elements (and most notably perhaps for Cr).  This implies that the residuals for these elements are correlated with, and thereby discriminative of, mean birth location in the disk.

\subsection{The holistic view of residual scatter in the Milky Way disk and halo}

\begin{figure*}[]
\centering
\includegraphics[scale=0.34]{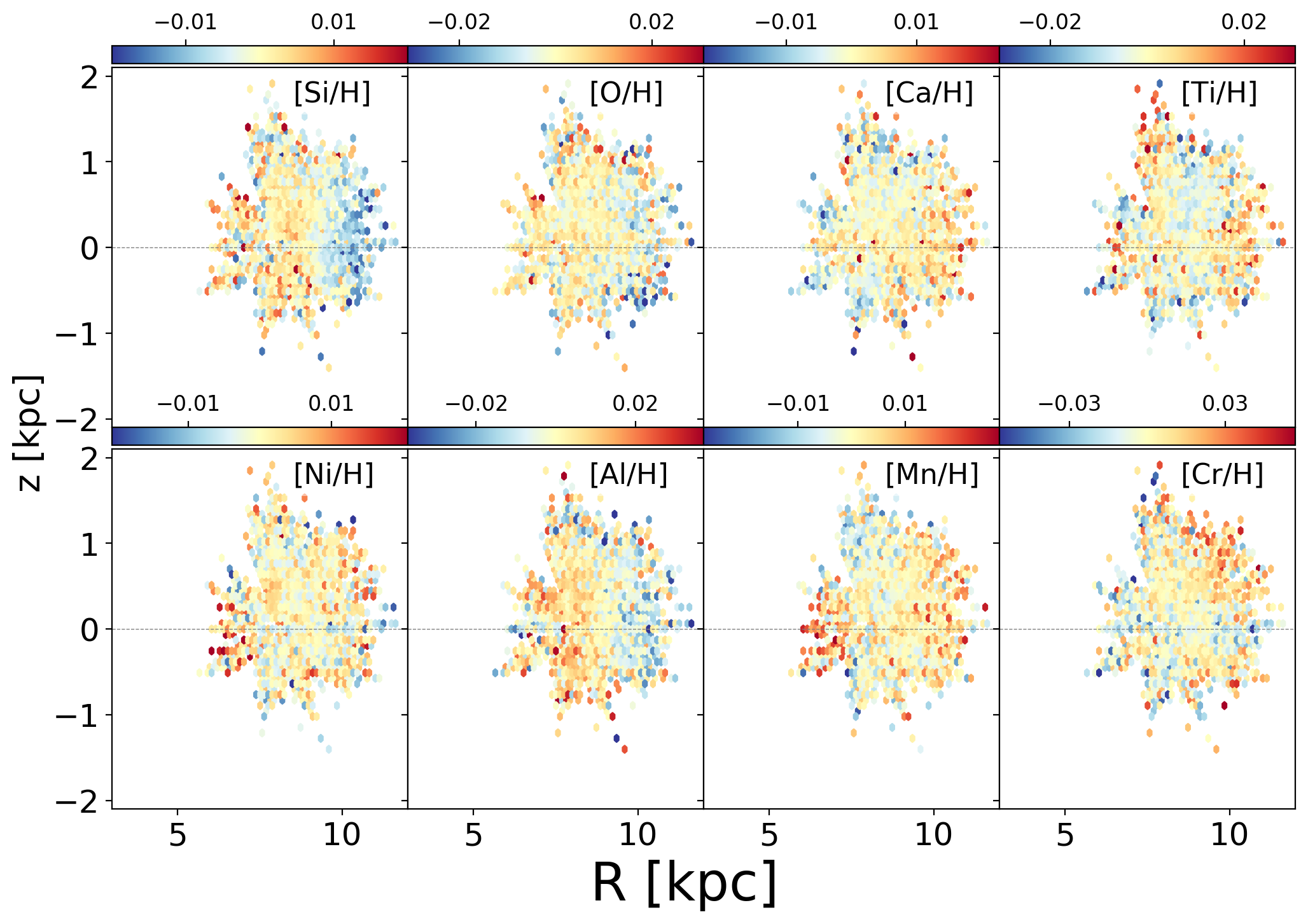}
\includegraphics[scale=0.34]{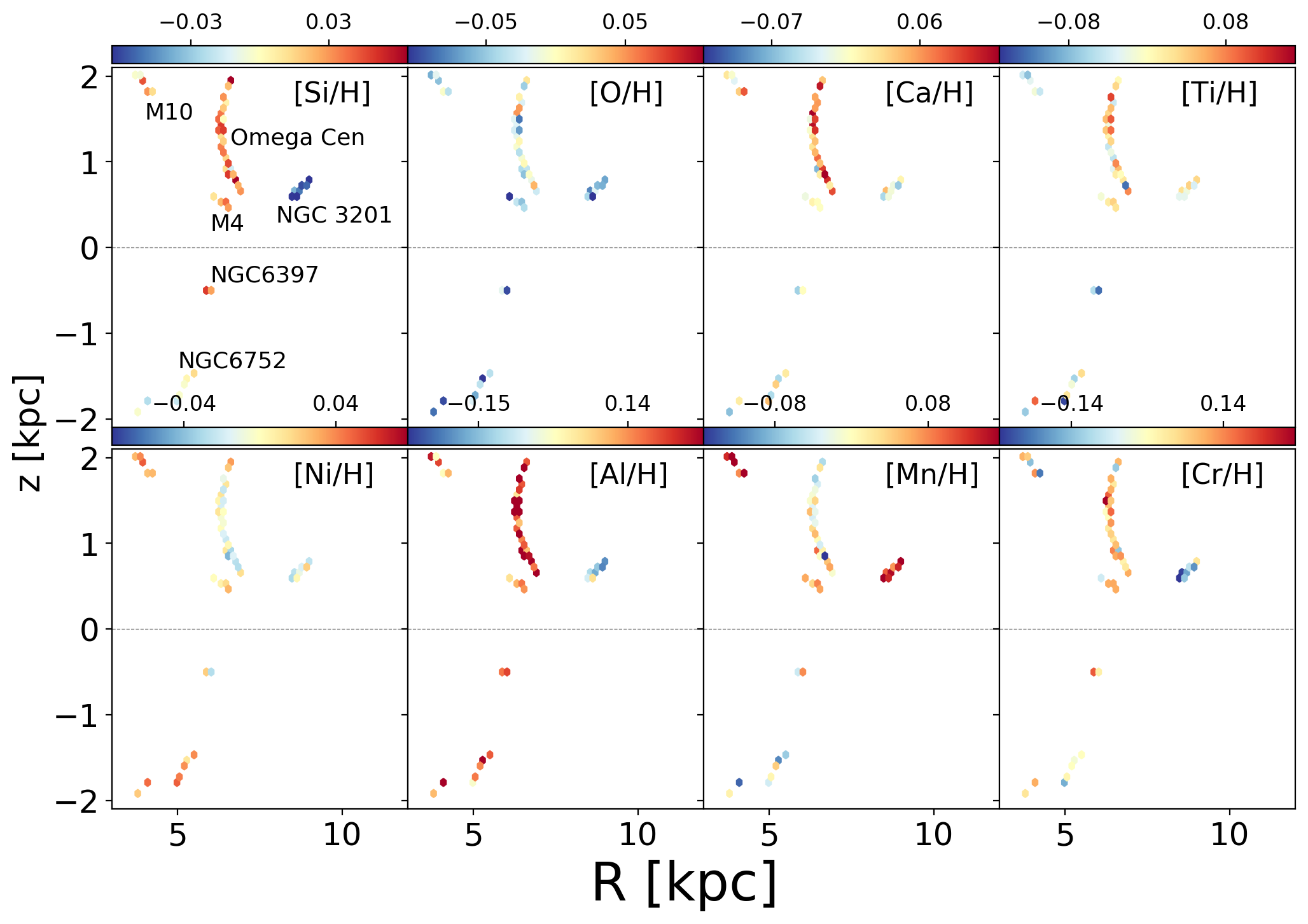}
\caption{The $R-z$ plane of disk stars with $\feh  > -1.0$~dex at left, and halo stars with $\feh < -1.0$~dex at right, colored by the mean amplitude of the element's residual in spatial bins. Note that the scale is different in the disk compared to the halo for the colorbar; the halo colorbars have a significantly larger extent. A minimum of $>$ 5 stars are shown in each bin. At left, the disk shows smooth structure in some residuals (i.e. for Si across radius). At right, overdensities in the observations light up with similar mean residual amplitudes at larger amplitudes than in the disk (and note that Al and Mn are anti-correlated for some of these structures). Presumably, in the halo, stars of different birth environments will group with common residual amplitudes, and/or the variance of the residuals within known groups may trace progenitor properties. The groups in the right hand panel that appear with the same mean residuals are systems that are targeted by APOGEE, as indicated in the [Si/H] panel of the halo. For example, the long streak of stars at $z = 1-2$~kpc is Omega Centauri tidal debris with large positive [Al/H] residuals. } 
\label{fig:halo}
\end{figure*}

We now demonstrate the potential of the information expressed in the element residuals from the prediction using (Fe, Mg) elements alone, across the Milky Way, more broadly. We proceed by comparing the residual scatter in the set of eight elements for disk stars to halo stars. The stars are selected to have SNR $>$ 100, and disk and halo separated only via a metallicity criteria; stars more metal rich than $\feh > -1.0$ are assigned as disk ($\sim$~80,000 stars) and stars more metal poor than $\feh < -1.0$~dex are assigned as halo ($\sim$3000 stars). Figure \ref{fig:halo} shows the spatial plane of Galactocentric radius and height $R-z$, for stars of the disk at left, and halo at right, with at least $>$ 5 stars per bin. Each bin in the $R-z$ plane is colored by mean residual amplitude, as shown in each Figure's colorbar. The colorbar in each panel is scaled to the 1-$\sigma$ standard deviation of the element's residual). The disk is smooth in residual trends, with some gradients seen for the elements radially, as explored in Figure \ref{fig:three}, in particular seen in Si. In the halo, the residual scatter is two or more times larger than in the disk, and groups of stars in the spatial plane have common and distinguishing mean residual amplitudes (but also in some cases significant scatter within individual sub-structure). Figure \ref{fig:halo} at right shows the stars that are grouped in spatial location have different mean residual amplitudes. The targeted APOGEE halo fields are labeled, and these correspond to the groups seen on the sky with similar residual amplitudes. Note the modeling of the halo and the residual amplitude is also sensitive to the neighbourhood $k$; for Figure \ref{fig:halo}, the $k$ is the same as the disk, of $k$=100. A full exploration of this is beyond the scope of the paper, but different $k$ have different utility in examining halo substructure.

\begin{figure*}[]
\centering
\includegraphics[scale=0.6]{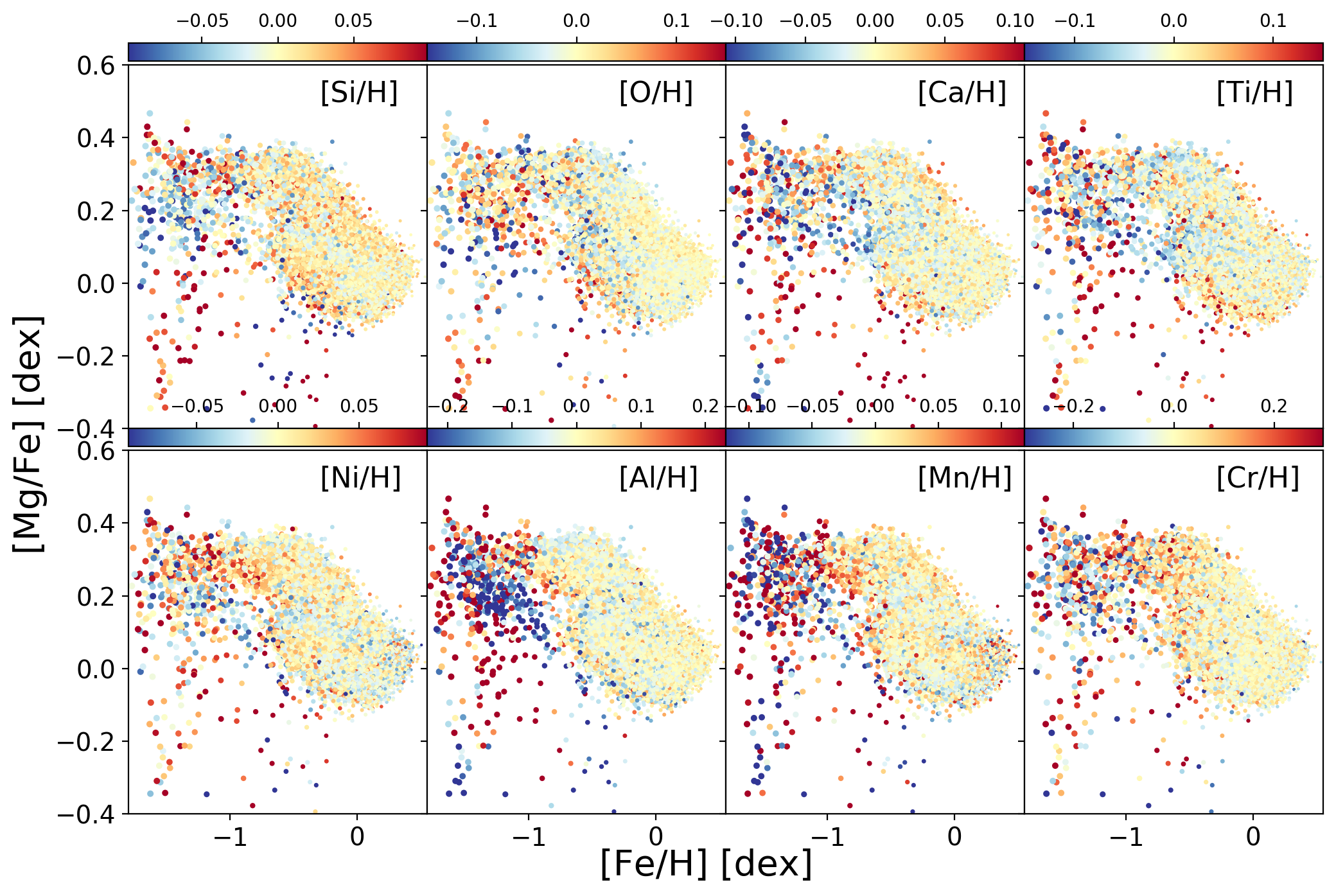}
\caption{The [Fe/H]-[$\alpha$/Fe] plane for 36,000 stars with SNR $>$ 200 including 500 stars with  $\feh < -1.0$~dex. The colorbar shows the abundance of the residual pertaining to each element indicated. The disk shows smooth structure across this plane whereas the more metal-poor stars are highly structured, likely as a consequence of the diversity of their birth environments. The points are scaled in size to be larger at decreasing [Fe/H].} 
\label{fig:halo2}
\end{figure*}

Figure \ref{fig:halo2} shows $\sim$36,000 stars selected with SNR $>$ 200, including $\sim$500 stars with $\feh < -1.0$~dex (with the same \teff\ and \logg\ cuts as in Section 2). The stars are coloured in the [Fe/H]-[Mg/Fe] plane by their residuals (the measured [X/H] $-$ the predicted [X/H]). This demonstrates the striking structure in the residuals in the population that is \textit{not} disk material. Presumably, the halo shows this interesting residual variability and structure as it is comprised of stars from many different star formation environments \citep[e.g. ][]{halo1_2021, halo2_2021, halo3_2020, halo4_2020, halo5_2021, Naidu2020, Buder2022}. Note where some elements show large positive residuals, others show large negative residuals. The elements Mn and Al are particularly powerful diagnostics of and within the halo sub-systems, compared to the disk; stars with negative residuals in [Al/H] look to be a potential additional chemical tag of the Gaia-Enceladus-Sausage system \citep[][]{Helmi2018, V2018}. This is not necessarily surprising given as Al and Mn have been used as diagnostic of halo substructure previously \citep[e.g.][]{Horta2021, Das2020}. Using the residuals effectively maps individual [X/H] measurements to a common chemical reference frame. Additionally, stars with low-[Mg/Fe] at higher metallicities, $\feh > -1.0$~dex show high residual amplitudes; a probable signature of these stars having an origin ex-situ the Milky Way disk.

\section{Discussion}

We demonstrated that using two abundances (Fe, Mg) for disk stars, we can predict an ensemble of other elements produced in supernovae to high precision, with $\sigma_{intrinsic} \approx 0.021$~dex. This reveals the (small) amplitude of the individuality of these elements, near the limits of the precision of many large surveys \citep[e.g.][]{GP2015, Kollmeier2017, Buder2018, Jofre2019}. The stars we examine comprise different metallicities, ages and birth radius, having migrated across the disk over time \citep[e.g.][]{Bensby2014, Minchev2018, Frankel2018, Frankel2019, Ratcliffe2021, Sharma2021}. From a galaxy evolution perspective, the extent to which (Fe, Mg) do not predict the other eight elements we study is, in part, due to environmental differences at birth in the radial and temporal extent of the disk, as told by the level of individuality of the elements \citep[see also][]{Ting2021, Weinberg2021}. However, we note that the correlations with present day radius and stellar age are small and that the most dominant correlations for some elements are imputed by artifacts in the spectra. We propose that the residual scatter we measure, after accounting for measurement uncertainties and astrophysical correlations,  affords a way of assessing the intrinsic scatter of elements at fixed birth radius and time in the disk. We note however, that the amplitude of the intrinsic scatter that we report for the elements is sensitive to the measurement errors, and this is the biggest source of uncertainty in our estimate. 

We explore correlations between the residuals of the measured [X/H] compared to the model's prediction from (Fe, Mg), as shown in Figures \ref{fig:three} and \ref{fig:four}. Looking in detail at these correlations, we see different behaviours for each element. We note that as the correlations between elements themselves are contaminated by systematic imprints (e.g. seen in Figure \ref{fig:four} and McKinnon et al., in preparation) we focus on only these age and orbital correlation, and not inter-element correlations.  A positive residual means that there is excess [X/H] compared to the model's prediction and a negative value means there is less than the prediction from (Fe, Mg) alone. The iron-peak elements of Mn and Ni show the largest age correlations, with positive residuals at youngest ages and negative residuals for old stars. This is presumably as younger stars are born in environments that have relatively higher contributions of SNIa. The primarily massive star, SNII elements, show (very small) correlations with mean guiding radius, measured with $L_z$. The elements Ca and Si correlate in opposite directions with $L_z$. The Si (explosive element) abundance is higher than the Ca (hydrostatic) element abundance at smaller guiding radius \citep[see also][]{Blancato2019}, yet their age relations are relatively flat. Regardless of age, this hints at a very small difference in the initial mass distribution of stars in the inner and outer parts of the Galaxy \citep[e.g.][]{Griffith2021, Krum2014, Matt2021}. 
  Quantitative modeling, however, is needed to understand and explain the underlying parameters of the environment that give rise to the systematic changes in individual abundance distributions over time and radius at fixed (Fe, Mg) as seen in the distributions of the residuals of these elements \citep[e.g.][and references therein]{Johnson2021, Spitoni2021, Matt2021, Philcox2019, Jan2017}.

  Overall, Figures \ref{fig:three} and \ref{fig:four} reveal that present day orbits and age explain only a fraction of the intrinsic scatter. After removing the astrophysical correlations of orbit and age with individual element residuals and accounting for errors, a level of $\sigma_{\mbox{intrinsic}} \approx 0.01-0.015$~dex in residual scatter around the model's prediction remains. This is a likely a measure of the scatter in the elements that are produced in supernovae, at fixed birth radius and age. Correspondingly, this is a complementary result to studies that use open clusters -- stars known to be born together, which report an intrinsic scatter of  $\leq 0.02-0.05$~dex for these elements within clusters \citep[e.g.][]{Bovy2016, Ness2018, Liu2019, Souto2019,  PJ2018, Cheng2021, Kos2021}. 
  
  We examine the residual distributions of the eight elements, predicted using (Fe, Mg, \logg, \teff) in open cluster systems. We find that these sites, comprising stars of common birth origin, have similar residual intrinsic dispersion measurements to the field, on average. Furthermore, that there are trends between the amplitude of the intrinsic dispersion of the residuals and [Fe/H], as seen in the field, and reported in Figure \ref{fig:one_summary}. However, the clusters show bias with respect to the field. That is, non-zero mean residual values. Some of this bias is consistent with correlations between the residual amplitudes and the orbital actions and age, as seen in the field. However, some correlations in the cluster population are not observed in the field population. The significant trend of the residual of Cr and the orbital action $J_z$, that is seen for the cluster stars and not  in the field, implies that the cluster dissolution process, heating and/or radial migration removes this relationship. Using the cluster sample, the residual for Cr is revealed to be a likely tag of mean birth $J_z$. Presumably, in general, we see stronger correlations with orbital actions and the cluster stars compared to the field, as the open clusters have not migrated significantly nor dissolved \citep[][]{Spina2020}.

  Although the residuals of the abundances at fixed (Fe, Mg) have very small amplitudes in the disk, the variance in the halo is substantially larger. Therefore in the halo, residual abundances may be useful to constrain and differentiate star formation environments and progenitors. Figure \ref{fig:halo} also reveals the large residual variance at very low [Mg/Fe]  ($\feh > -1.0$~dex, $\mgfe < -0.2$~dex), which is presumably ex-situ material. In the disk, where the intrinsic variance is very small, stars that are significant outliers from the model's prediction may be interesting objects to study. These may have had their abundances modified over the course of their evolution, due to dynamical effects like rotation, binarity and planet accretion.

\section{Conclusions}

The detailed star formation environment across the Galactic disk over time is traced by the variance in element abundances, [X/H], at fixed fiducial supernovae contributions of (Fe, Mg).
We quantify this variance by building local linear models to predict eight abundance ratios [X/H] (for X = Si, O, Ca, Ti, Ni, Al, Mn, Cr)  that are produced in supernovae, given the chemical abundances of [Fe/H] and [Mg/H] and evolutionary state parameters of  \logg\ and \teff. Investigation of correlations between the residual of the model (measured [X/H] $-$ predicted [X/H]) and astrophysical parameters reveals that the residuals in part correlate with orbit and age. However, the residual measurement, which is near the measurement precision limits, is also hampered by nuisance signals in the spectra, traced by radial velocity correlations. These signals will be important to model and remove in order to fully exploit the information content in stellar spectra \citep[e.g.][]{Feeney2021, DeM2021, Wheeler2021, PJ2018}. Overall, we find that (Fe, Mg) predict the other abundances with a median intrinsic dispersion of $\approx 0.021$~dex, which decreases to  $\approx 0.15$~dex once additional non-chemical parameters are added to the model. 

We claim, based on this study that disk stars in the Milky Way have $\approx 0.01-0.015$~dex, on average, scatter in elements synthesised in supernovae, at fixed birth radius and time. This corresponds to $\sim 20-50$~percent of the individual element variance at fixed (Fe, Mg), for elements produced in supernovae. We report some weak dependence of the intrinsic scatter of the residuals of each element on [Fe/H]. The intrinsic scatter of individual elements, at fixed birth radius and time, plus any [Fe/H] dependence, has implications for initial cluster mass distributions and timescales of formation (which are presumably similar regardless of metallicity and age), as well as the mechanics of element mixing  \citep[e.g.][]{BH2010, Krumholz2019, arm2018}.
Chemical evolution modeling to fit these data will inform the environmental parameters that give rise to the trends we see in individual elements across both present-day radius and age \citep[see][and references therein]{Matt2021}. We expect new survey data will enable us to expand this assessment of the intrinsic scatter at fixed radius and time in the disk, to examine different elements and nucleosynthetic families \citep[e.g.][]{deSilva2015, Buder2021, Kollmeier2017}. 

Finally, we highlight that this study has focused on the  disk population ($\feh > -1.0$~dex), where the joint element abundance variance is marginal \citep[][]{Feeney2021, Rampalli2021, Lu2021, Ness2019, Weinberg2019}. This is not the case in the halo ($\feh < -1.0$~dex) where residual abundances calculated here are likely much stronger tags of common birth environment.

\section{Acknowledgements}

Melissa K. Ness acknowledges support from a Sloan Foundation Fellowship. We thank the dynamics group at Flatiron for helpful discussions with respect to this work during its development. 

\section{Appendix}

\renewcommand{\thefigure}{A\arabic{figure}}
\setcounter{figure}{0}

\begin{figure}
\centering
\includegraphics[scale=0.31]{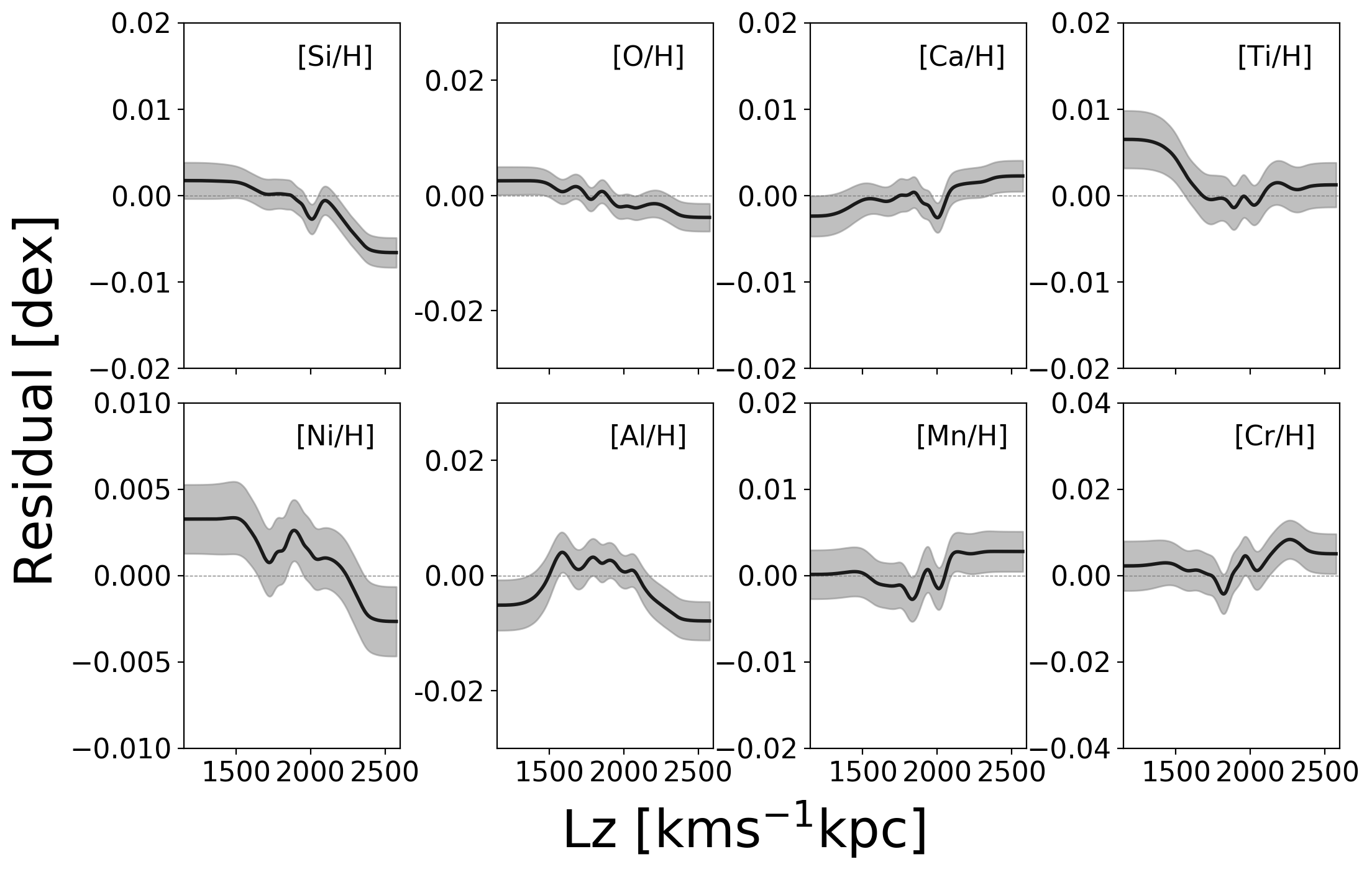}\\
\includegraphics[scale=0.31]{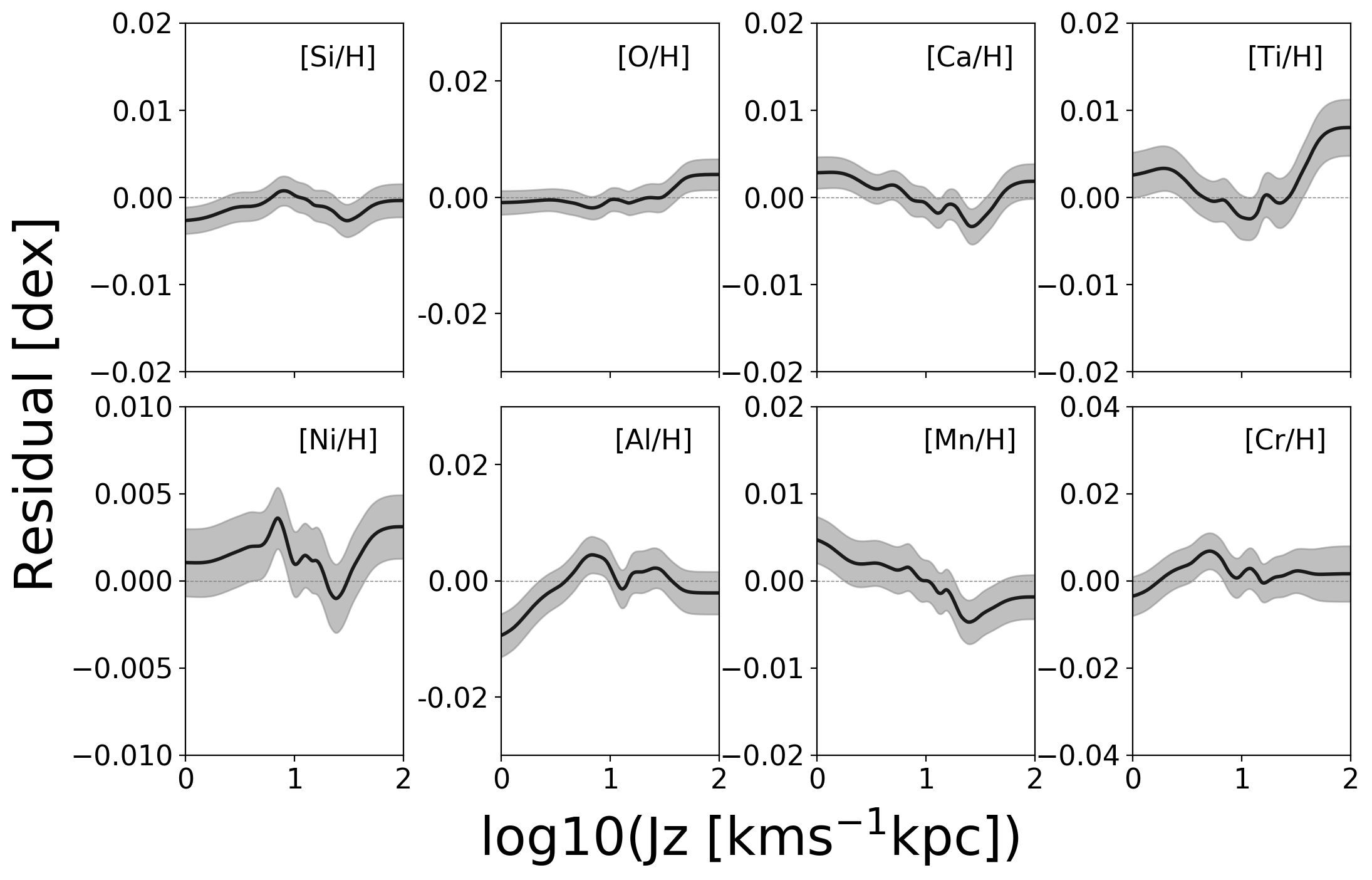}
\caption{The running mean of the angular momentum $L_z$ at top, and the log of the vertical action $J_z$ at bottom, versus the residual amplitude for each element for stars with ages $<$ 3.5 Gyr, with the y-axis for each scaled to the intrinsic dispersion. The astrophysical correlation with $L_z$ represents only a fraction of the residual amplitude but the trends are stronger than that reported in Figures \ref{fig:three} and \ref{fig:four}. The shaded region represents the confidence on the mean.} 
\label{fig:three_appendix}
\end{figure}

Figure \ref{fig:three_appendix} shows the $\sim$~3300 stars with ages $ < 3.5$~Gyr and the running mean of their residuals in the eight elements and $L_z$ and $J_z$. These trends between residuals of the elements and orbital actions are stronger for the younger population compared to the full field distribution.

\clearpage

\bibliography{mknbib}
\end{document}